\documentclass[notitlepage,prd,showkeys,twocolumn,preprintnumbers,floatfix,
superscriptaddress,nofootinbib]{revtex4-1}
\usepackage{amsmath}
\usepackage{amssymb}
\usepackage{xcolor}
\usepackage{bbm}
\usepackage{braket}
\usepackage{xspace}
\usepackage{mathtools}
\usepackage{graphicx}
\usepackage{enumerate}
\usepackage{hyperref}
\usepackage{verbatim}
\usepackage{tabularx}
\usepackage{multirow}
\usepackage{enumitem}
\usepackage{orcidlink}
\usepackage{nicefrac}

\usepackage{mfirstuc}
\newcommand{\addReviewer}[2]{
\expandafter\newcommand\csname #1\endcsname[1]{{\sf \color{#2} {#1}:\,##1}}
\expandafter\newcommand\csname #1cor\endcsname[2]{{\color{#2}
{#1}:\,\st{##1}{\sf ##2}}}
\expandafter\newcommand\csname #1color\endcsname{#2}
}

\usepackage{soul,color}
\definecolor{chromeyellow}{rgb}{1.0, 0.65, 0.0}
\definecolor{DodgeBlue}{rgb}{0.118, 0.565,1.000}
\definecolor{asparagus}{rgb}{0.53, 0.66, 0.42}
\definecolor{cadmiumgreen}{rgb}{0.0, 0.42, 0.24}

\definecolor{jlab_red}{RGB}{192,39,45}
\definecolor{jlab_orange}{RGB}{249,102,0}
\definecolor{jlab_blue}{RGB}{47,122,121}
\definecolor{jlab_green}{RGB}{65,125,10}

\addReviewer{MH}{DodgeBlue}
\addReviewer{MB}{jlab_red}
\addReviewer{FB}{jlab_blue}

\hypersetup{
pdftitle = {spectral-finite-volume},
pdfsubject = {Finite-volume spectral densities},
pdfkeywords = {Spectral densities, QCD, Hadron, Physics, Lattice},
pdfauthor = {Bresciani, Bruno, Hansen},
pdfnewwindow = {true},
colorlinks = {true},
linkcolor = {jlab_blue},
citecolor = {jlab_blue},
filecolor = {jlab_blue},
urlcolor = {jlab_blue}
}

\newcommand{\milan}{Dipartimento di Fisica, Universit\`a degli Studi di
Milano-Bicocca, Piazza della Scienza 3, I-20126 Milano, Italy}
\newcommand{\infn}{INFN, Sezione di Milano-Bicocca, Piazza della Scienza 3,
I-20126 Milano, Italy}
\newcommand{\edinburgh}{Higgs Centre for Theoretical Physics, School of Physics
and Astronomy, The University of Edinburgh, Edinburgh EH9 3FD, UK}

\begin{document}

\title{Finite-volume effects on smeared spectral densities}

\author{Francesca A.~Bresciani~\orcidlink{0009-0009-3299-5819}}
\email[e-mail: ]{f.bresciani4@campus.unimib.it}
\affiliation{\milan}
\affiliation{\infn}

\author{Mattia~Bruno~\orcidlink{0000-0002-5127-4461}}
\email[e-mail: ]{mattia.bruno@unimib.it}
\affiliation{\milan}
\affiliation{\infn}

\author{Maxwell T.~Hansen~\orcidlink{0000-0001-9184-8354}}
\email[e-mail: ]{maxwell.hansen@ed.ac.uk}
\affiliation{\edinburgh}

\begin{abstract}
Using two distinct approaches, we derive a universal expression for the leading
finite-volume effects of the smeared vector-vector spectral density
(proportional to the smeared hadronic $R$-ratio) in a periodic cubic spatial
volume of side length~$L$. First, building on the results of previous work for
finite-volume effects on Euclidean two-point functions, we show that the $L$
dependence is exponentially suppressed for a certain class of smearing kernels,
and that the leading effects can be expressed universally in terms of the pion
form factor. The same representation is then derived starting from the
Lellouch-L\"uscher-Meyer expression for the spectral decomposition of the
correlator. The results may prove useful for controlling the $L \to \infty$
extrapolation of smeared spectral densities, in particular by defining a
scaling regime in which the finite-volume effects are dominated by the leading
terms in a large $L$ expansion and thus can be reliably estimated. To
illustrate this point, we also present numerical estimates based on various kernels and models of particle interactions.
Despite focusing on the vector channel, our derivation defines a general framework applicable to other cases as well.
\end{abstract}
\date{\today}

\maketitle
% {\color{teal} \textbf{V2 proposed modifications - TO DO}:
% \begin{itemize}
    % DONE (9 Oct 2026): real axis extended to $[-5,5]$, label changed to 'pion branch point', markers by type with legend; see figs/make_final_integrand_fig.py
    % \item \textbf{Update FIG. 3}: extend the real axis to $[-5,5]$; explicitly show the sqrt branch cut along the imaginary axis OR just rename 'pion pole' as 'pion branch cut' in the label (or similar).
%
%     \item \textbf{Numerical section}, second-to-last paragraph before the conclusions: add the equation for the contribution from the residues at the Cauchy poles in the free case (exponentially suppressed envelope + oscillatory term). Adjust the text accordingly.
%
%     \item \textbf{Appendix A}: further work could be done on the term $\delta G^B$ and its contribution to the smeared spectral density, and consequently on our analytical prediction. Key observation: Eq. (2.37) of Max \& Agostino's paper shows that $\delta G^B$ is exponentially decaying in $x_0 (\equiv \tau)$ as well, so our logic can in principle be applied.
% \end{itemize}
% }

\section{Introduction}

Quantum field theory describes a rich range of physical systems, from the
interactions of fundamental particles to various condensed matter and other
material science applications. Interesting phenomenology is particularly visible in the
non-perturbative regime, in which a series expansion about the non-interacting
limit does not provide a useful description of the true dynamics. In this case,
reliable analytic methods are often unavailable.

One approach that has proven highly successful is to employ numerical lattice
quantum field theory, i.e.~to evaluate the Euclidean quantum path integral in a
finite discretized spacetime using Monte Carlo importance sampling. This has
been particularly impactful in the context of numerical lattice quantum
chromodynamics (QCD), where it is applied to the strongly interacting sector of
the Standard Model of Particle Physics, allowing for percent-level,
first-principles predictions of quantities such as the proton and neutron
masses~\cite{BMW:2014pzb} and the leading hadronic effects on the magnetic
moment of the muon~\cite{Aliberti:2025beg}.

A general limitation of this strategy is that it yields discrete samples of
finite-volume Euclidean (equivalently, imaginary-time) correlators with
statistical uncertainties. Thus, although the Osterwalder-Schrader
theorem~\cite{Osterwalder:1973dx} generally guarantees that it is possible to
continue from imaginary to real time with complete analytic knowledge, the
formal result is of limited utility with realistic numerical data. In many
cases, this issue can be circumvented by directly relating the Euclidean
correlation function to an observable of interest, e.g.~by fitting to a single
exponential at asymptotic separations in order to extract the ground state of
the system for a particular set of quantum numbers. Recently, the community has
seen a resurgence of interest in the alternative approach of instead relating
the Euclidean correlator to a smeared spectral density.

For concreteness, we focus here on the vector-vector Euclidean two-point
correlator in a finite cubic spatial volume of side length $L$ with periodic
boundary conditions, projected to zero spatial momentum,\footnote{Generically
lattice QCD calculations are performed in a finite temporal extent as well, but
we assume that this is sufficiently large that the effects of the temporal
boundaries are negligible.}
\begin{equation}
G_L(\tau) = -\frac{1}{3} \sum_{k=1}^{3} \int_{0}^{L} {\rm d}^3 \boldsymbol x \,
\langle j_k(x) j_k(0) \rangle_{L} \,,
\label{eq:GL-vv-def}
\end{equation}
where $j_\mu(x)$ is the Euclidean-signature isospin-triplet vector current, and
$x = (\boldsymbol x, \tau)$ is the Euclidean coordinate. The associated
finite-volume spectral density is defined as
\begin{equation}
\rho_L(\omega) = -\frac{L^3}{3 \omega^2} \sum_{k=1}^{3} \langle 0 \vert j_k(0)
\delta_{\widehat {\boldsymbol P}, \boldsymbol 0} \delta(\omega - \widehat H)
j_k(0) \vert 0 \rangle_L \,,
\label{eq:rhoL-def}
\end{equation}
where $\widehat H$ and $\widehat {\boldsymbol P}$ are the Hamiltonian and
momentum operators, respectively. Here the Kronecker delta
\begin{equation}
L^3 \delta_{\widehat {\boldsymbol P}, \boldsymbol 0} = \int_{0}^{L} {\rm d}^3
\boldsymbol x \, e^{i \widehat {\boldsymbol P} \cdot \boldsymbol x} \,,
\end{equation}
ensures that only zero-momentum states contribute to the spectral density.

The objects defined in Eqs.~\eqref{eq:GL-vv-def} and \eqref{eq:rhoL-def} are
related via a Laplace transform:
\begin{equation}
G_L(\tau) = \int_0^{\infty} {\rm d}\omega \, \omega^2 \, e^{-\omega \tau} \,
\rho_L(\omega) \,.
\label{eq:GL-rhoL}
\end{equation}
A generalization of this relation is achieved by noting that $\omega^2 \,
e^{-\omega \tau}$ can be replaced with a generic smooth function, leading to
the definition of the smeared spectral density,
\begin{equation}
\widehat{\rho}_{L,\kappa}(\boldsymbol{\alpha}) = \int_0^{\infty} {\rm d}\omega
\, \widehat{\kappa}(\omega, \boldsymbol \alpha)\, \rho_L(\omega) \,,
\label{eq:rho-hat-def}
\end{equation}
where $\widehat{\kappa}(\omega, \boldsymbol{\alpha})$ represents a smearing
kernel with shape parameters collected in the vector $\boldsymbol{\alpha}$.

Taking inspiration from the Backus-Gilbert reconstruction
method~\cite{Backus:1968svk} as well as various recent developments in the
context of lattice
QCD~\cite{Meyer:2007fc,Meyer:2007ic,Meyer:2007dy,Meyer:2008gt,Brandt:2012jc,
Hansen:2017mnd,Hansen:2019idp,Bruno:2020kyl,Horak:2021syv,DelDebbio:2021whr,
Pawlowski:2022zhh,Frezzotti:2023nun,Horak:2023xfb,Bruno:2024fqc,
DelDebbio:2024lwm,Candido:2024hjt,Dutrieux:2024rem,Frezzotti:2025hif,
Dutrieux:2025jed,Lupo:2026vdj,Tsuji:2026zku}, we focus here on linear
reconstruction of spectral functions from Euclidean correlators.%
\footnote{Non-linear methods for reconstructing spectral densities have also
been thoroughly investigated, but are not the focus of this work. See
e.g.~Refs.~\cite{Burnier:2013nla,Fei:2021,Huang:2023gpb,Bergamaschi:2023xzx,
Fields:2025glg} as well as the review Ref.~\cite{Rothkopf:2022fyo}.}
Suppose that
$G_L(\tau)$ is sampled at values $\tau_i$ for $i \in \{1, 2, \cdots, N\}$. One
can then determine a set of coefficients $c_{i}(\boldsymbol{\alpha})$, designed
so that
\begin{equation}
\widehat{\kappa}_{\sf r}(\omega \vert \boldsymbol c( \boldsymbol{\alpha})) =
\omega^2 \sum_{i=1}^{N} c_{i}(\boldsymbol{\alpha}) \, e^{-\omega \tau_i} \,,
\label{eq:kappa-expansion}
\end{equation}
approximates $\widehat{\kappa}(\omega, \boldsymbol{\alpha})$, where the
subscript ${\sf r}$ denotes the reconstructed version.
Then $\widehat{\rho}_{L,\kappa}$ is estimated by applying the same coefficients
to $G_L$. We denote the approximated version by $\widehat{\rho}_{L,\kappa, {\sf
r}}$:
\begin{equation}
\widehat{\rho}_{L,\kappa, {\sf r}}(\boldsymbol{c}(\boldsymbol{\alpha})) =
\sum_{i=1}^{N} c_{i}(\boldsymbol{\alpha}) \, G_L(\tau_i) \,.
\label{eq:G-to-rho-hat}
\end{equation}
This construction generally depends on a regularization procedure, which we do
not specify here.
If one sends $N \to \infty$ and removes the regulator on
Eq.~\eqref{eq:G-to-rho-hat} then $\widehat{\kappa}_{\sf r} \to
\widehat{\kappa}$ and $\widehat{\rho}_{L,\kappa, {\sf r}} \to
\widehat{\rho}_{L,\kappa}$~\cite{Backus:1968svk,Hansen:2019idp,Bruno:2024fqc}.

In practice, this formal $N \to \infty$ limit cannot be achieved with realistic
lattice data. The goal is instead to quantify the systematic uncertainty
associated with the approximation of $\widehat{\kappa}$ by
$\widehat{\kappa}_{\sf r}$, and to choose $c_i (\boldsymbol{\alpha})$ such that
this is subdominant to the statistical uncertainty on $\widehat
{\rho}_{L,\kappa, {\sf r}}$. Denoting the covariance of $G_L(\tau_i)$ by
$\Sigma_{ij}$, the statistical variance on the smeared spectral function is
\begin{align}
\sigma_{\sf stat}^2 & = \sum_{i,j=1}^{N} c_i(\boldsymbol{\alpha}) \Sigma_{ij}
c_j(\boldsymbol{\alpha}) \,;
\label{eq:stat-variance}
\end{align}
and a possible definition of the systematic uncertainty due to the
approximation of $\widehat{\kappa}$ by $\widehat{\kappa}_{\sf r}$ is
\begin{align}
\begin{split}
\sigma_{\sf sys} & = \left \vert \int_0^{\infty} {\rm d}\omega \, \Big [
\widehat{\kappa}(\omega, \boldsymbol{\alpha}) - \widehat{\kappa}_{\sf r}(\omega
\vert {\boldsymbol c}(\boldsymbol{\alpha}) ) \Big ] \rho_L(\omega) \right \vert
\,, \\
& = \Big \vert \widehat {\rho}_{L,\kappa}(\boldsymbol{\alpha}) -
\widehat{\rho}_{L,\kappa, {\sf r}}({\boldsymbol c}(\boldsymbol{\alpha})) \Big
\vert
\,.
\label{eq:sys-variance}
\end{split}
\end{align}
The optimal solution is then achieved when $\sigma_{\sf stat} \approx
\sigma_{\sf sys} $, with a possible multiplicative factor to ensure that the
systematic uncertainty is subdominant.

While $\sigma_{\sf stat}$ can be directly estimated from the data, $\sigma_{\sf
sys}$ depends on the true spectral density $\rho_L(\omega)$ and is therefore
unknown. As a result, one must consider different strategies to either model or
bound $\sigma_{\sf sys}$. For example, one can perform a sensitivity analysis
by varying the model parameters and observing how the reconstructed smeared
spectral density changes, as in Ref.~\cite{Hansen:2019idp}.

Within lattice QCD, the approach of spectral reconstruction has seen increased
attention, particularly in the context of zero-temperature
calculations~\cite{Bulava:2021fre,Lupo:2021nzv,
ExtendedTwistedMassCollaborationETMC:2022sta,DeSantis:2022qht,
DelDebbio:2022qgu,Panero:2023zdr}.
The essential idea is to first reliably extract $\widehat{\rho}_{L, \kappa}$
with both systematic and statistical uncertainties, and then to estimate the $L
\to \infty$ limit to reach an infinite-volume smeared spectral density, denoted
$\widehat{\rho}_{\kappa}$. One can subsequently remove the smearing width,
possibly also taking a chiral and continuum limit, to obtain a full physical
prediction for a given spectral quantity. For other applications it may be
useful to restrict to results at fixed smearing width, which can themselves be
directly compared to experiment~\cite{ExtendedTwistedMassCollaborationETMC:2022sta,Juttner:2026fui}.

In this work, we are concerned with improving the theoretical control of the $L
\to \infty$ extrapolation at fixed smearing kernel $\widehat{\kappa}(\omega,
\boldsymbol{\alpha})$. In particular, we write
\begin{equation}
\widehat{\rho}_{L, \kappa}(\boldsymbol{\alpha}) =
\widehat{\rho}_{\kappa}(\boldsymbol{\alpha}) + \delta \widehat{\rho}_{L,
\kappa}(\boldsymbol{\alpha}) \,,
\label{eq:rho-hat-decomposition}
\end{equation}
where $\widehat{\rho}_{\kappa}(\boldsymbol{\alpha}) = \lim_{L \to \infty}
\widehat{\rho}_{L,\kappa}(\boldsymbol{\alpha}) $ and thus $\delta
\widehat{\rho}_{L, \kappa}(\boldsymbol{\alpha})$ represents the finite-volume
correction that vanishes in the infinite-volume limit.

Our key message is that, for a certain class of smearing kernels,
$\delta \widehat{\rho}_{L, \kappa}$ admits a large-volume expansion with terms
that decay exponentially in $L$, generally with power-law prefactors.
This is the same situation as for the Euclidean correlator itself, is similar
to the result for stable hadron masses \cite{Luscher:1985dn}, and is generally
referred to as ``exponentially suppressed'' finite-volume effects. However, as
we stress in Sec.~\ref{sec:numerics}, such a representation is only
useful when the leading terms in the series dominate. Our main result,
Eq.~\eqref{eq:final-result} below, allows one to check convergence explicitly
for a given choice of smearing kernel and thereby identify the onset of the
asymptotic regime, in which the $L \to \infty$ extrapolation is under control.
In addition, the result may be used to either remove leading finite-volume
effects or to perform a more informed extrapolation by including such effects
in the $L$-dependent fit function.

We derive the main result in two different ways: First, we use the expression
for finite-volume effects on $G_L(\tau)$ derived in
Refs.~\cite{Hansen:2019rbh,Hansen:2020whp} and apply the coefficients
$c_i(\boldsymbol{\alpha})$ to directly obtain $\delta \widehat{\rho}_{L,
\kappa}$. Second, we start from the Lellouch-L\"uscher-Meyer formalism for
relating finite- and infinite-volume matrix elements and use that to directly
reach a representation for $\delta \widehat{\rho}_{L, \kappa}$ that is
equivalent to the first derivation. Certain approximations are made in both
derivations, and it is instructive to understand the regime of validity of the
final result from two different perspectives.

In the context of lattice QCD calculations of the anomalous magnetic moment of
the muon, both methods have been applied in practice to estimate finite-volume
effects and the results have been found to be numerically
consistent~\cite{Ce:2022kxy,Ce:2022uix,FermilabLatticeHPQCD:2023jof,
MILC:2024ryz,Djukanovic:2024cmq,Beltran:2026ofp,Beltran:2026ktg}. This work
therefore also gives a theoretical understanding of this consistency.

The remainder of this manuscript is organized as follows. In the next section,
we recall the expression for finite-volume effects on $G_L(\tau)$ as derived in
Refs.~\cite{Hansen:2019rbh,Hansen:2020whp}.
We then show how the integral expression can be converted, via a Wick rotation,
into a form with an integrand that is exponentially decaying in $\tau$, to
which one can apply Eq.~\eqref{eq:kappa-expansion} directly to reach a
$\widehat \kappa$-weighted integral expression for $\delta \widehat{\rho}_{L,
\kappa}$. The Wick rotation also has the consequence of transitioning from
spacelike to timelike kinematics, thereby clarifying how the finite-volume
effects on $G_L(\tau)$ can be expressed in terms of either. In
Sec.~\ref{sec:LL}, we then derive the same result using the Lellouch-L\"uscher
and Meyer formalism for connecting finite- and infinite-volume matrix elements.
In Sec.~\ref{sec:numerics}, we evaluate the result numerically for different
choices of the smearing kernel and the form factor. Finally, in
Sec.~\ref{sec:conclusions}, we briefly conclude and give an outlook for future
applications of these ideas. We include three appendices containing details of
the derivations.

\section{Derivation}\label{sec:derivation}

We begin with the finite-volume effects on $G_L(\tau)$, defined in
Eq.~\eqref{eq:GL-vv-def}. Defining $G(\tau) = \lim_{L \to \infty} G_L(\tau)$,
we are interested in the difference between the finite- and infinite-volume
correlators, which we denote by $\delta G_L(\tau)$:
\begin{equation}
\delta G_L(\tau) = G_L(\tau) - G(\tau) \,.
\label{eq:deltaGL-def}
\end{equation}

\begin{figure}[t]
\centering
\includegraphics[width=0.4\textwidth]{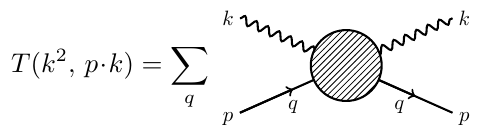}
\caption{Kinematics of the forward Compton amplitude appearing in the
finite-volume effects on the vector-vector correlator. The solid lines
represent pions with charge $q$ while the wavy lines represent external
currents.}
\label{fig:Compton}
\end{figure}

These were shown in Refs.~\cite{Hansen:2019rbh,Hansen:2020whp} using a generic
all-orders low-energy effective theory of pions to take the form\footnote{Up to
neglected terms falling as $e^{- \alpha m L}$ where $\alpha = \sqrt{2 +
\sqrt{3}} \approx 1.93$. These are further suppressed in practice, as they
require at least two-loop Feynman diagrams when evaluated in chiral
perturbation theory. See Ref.~\cite{Hansen:2020whp} for details.}
\begin{multline}
\delta G_L(\tau)= - \frac{1}{6} \sum_{\boldsymbol n \neq \boldsymbol 0} \int
\frac{{\rm d}p_3}{2 \pi} \frac{e^{-\vert \boldsymbol n \vert L \sqrt{m^2 +
p_3^2}}}{4 \pi \vert \boldsymbol n \vert L} \\ \times \int \frac{{\rm d}k_3}{2
\pi} \cos{(k_3 \tau)} \mathrm{Re} T(-k_3^2, -p_3 k_3)\,,
\label{eq:deltaGL-HP}
\end{multline}
where $m$ is the pion mass and $T$ is the forward Compton amplitude, with
kinematics as illustrated in Fig.~\ref{fig:Compton}. The latter can be
decomposed as $T = T_{\mathrm{pole}} + T_{\mathrm{reg}}$, where the pole
contribution is defined as
\begin{equation}
T_{\mathrm{pole}}(-k_3^2, - p_3 k_3) = \sum_{x = \pm} \frac{2 (4m^2 + k_3^2)
F(-k_3^2)^2}{k_3^2 + 2 x k_3 p_3 - i \epsilon} \,,
\label{eq:T-pole}
\end{equation}
where $F$ is the electromagnetic pion form factor, sampled in the spacelike
region when its argument is negative.

The decomposition of $T$ leads to a corresponding split of the finite-volume
effects,
\begin{equation}
\delta G_L(\tau) = \delta G_L^{\mathrm{pole}}(\tau) + \delta
G_L^{\mathrm{reg}}(\tau)\,.
\label{eq:deltaGL-pole-reg}
\end{equation}
As was described in Refs.~\cite{Hansen:2019rbh,Hansen:2020whp}, the pole
contribution dominates the finite-volume effects for realistic volumes and pion
masses, while the regular part is subdominant.

Restricting to the pole contribution and following
Refs.~\cite{Hansen:2019rbh,Hansen:2020whp}, we note that it can be decomposed
into two terms,
\begin{align}
\delta G_L^{\mathrm{pole}} (\tau)
& = \delta G_L^A(\tau) + \delta G_L^B(\tau) \,,
\label{eq:deltaGL-pole-AB}
\end{align}
where
\begin{multline}
\delta G_L^A(\tau)= \frac{1}{6} \sum_{\boldsymbol n \neq \boldsymbol 0}
\mathrm{Im}\int_{\mathbb{R}+i\mu} \frac{{\rm d}k_3}{2 \pi} \ \frac{e^{-\vert
\boldsymbol n \vert L \sqrt{m^2 + k_3^2/4}}}{4 \pi k_3 \vert \boldsymbol n
\vert L} \\ \times e^{ik_3 |\tau|} (4m^2 + k_3^2)F(- k_3^2)^2 \,,
\label{eq:deltaGL-pole-A}
\end{multline}
and $\delta G_L^B(\tau)$ is again subdominant for realistic volumes and for a
realistic functional form of the form factor, as discussed in
Appendix~\ref{app:B-term-subdominant}.

The integration contour in Eq.~\eqref{eq:deltaGL-pole-A} is shifted into the
complex plane by an amount $\mu$, with $0 < \mu \leq 2m$. The lower bound is
required to avoid the $1/k_3$ pole on the real axis,\footnote{Note in fact that
the integral is perfectly well-defined for $\mu=0$ as the integrand depends on
$\sin(k_3 \vert \tau \vert)/k_3$, which has a removable and thus integrable
singularity at $k_3=0$. However, this corresponds to averaging the $\mathbb R+i
\mu$ and $\mathbb R - i \mu$ results and is not equal to the integral defining
$\delta G_L^A(\tau)$.} while the upper bound ensures that one does not cross
the branch point of the square root function. As discussed in more detail
below, the form factor generally cannot contain poles that restrict the allowed
range of $\mu$.
An exception is the case of unphysically heavy pion masses, for which a
first-sheet pole develops, corresponding to a stable vector-meson bound state.

\begin{figure*}[t]
\centering
\includegraphics[width=\textwidth]{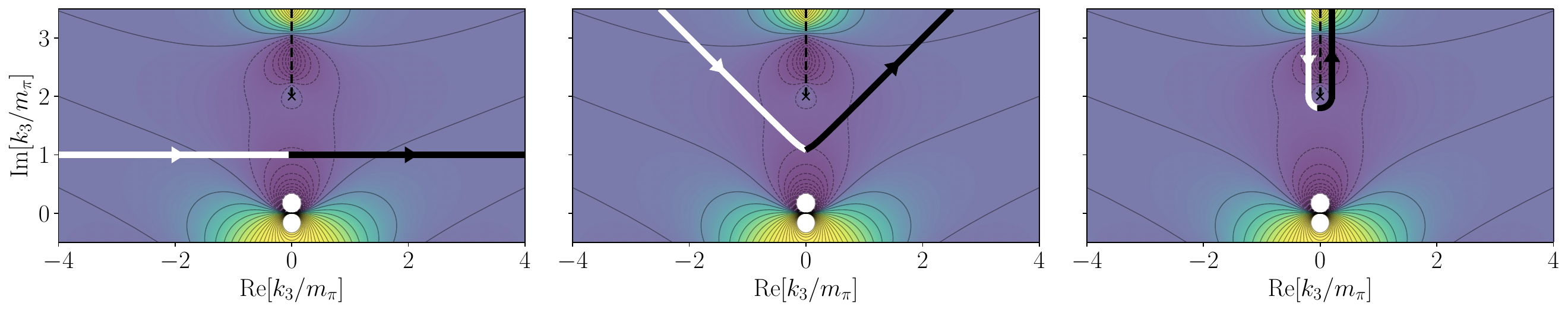}
\caption{Representation of the Wick rotation that relates the $\delta
G_L^A(\tau)$ integrand that is decaying in $L$ and oscillatory in $\tau$
[\emph{left panel}; corresponding to Eq.~\eqref{eq:deltaGL-pole-A}], to a form
that is decaying in $\tau$ and oscillatory in $L$ [\emph{right panel};
corresponding to Eq.~\eqref{eq:deltaGL-final}]. The middle panel illustrates an
intermediate step, where the contour is rotated by an angle $\theta$. The key
observation is that the black- and white-contoured integrals are equal for all
three panels, with the former generating $\delta G_L^{A,+}(\tau)$ and the latter
$\delta G_L^{A,-}(\tau)$. This follows from the analyticity of the integrand and
the fact that it falls off sufficiently rapidly at large radius. We also show
an example integrand as a contour plot in the complex $k_3$ plane, here for the
Gounaris-Sakurai form factor model (see Sec.~\ref{sec:numerics} for details),
with $mL = 4$ and $m\tau = 1$. The axis labels here and in
subsequent figures are written in terms of the pion mass
$m_\pi$, so that the figures are self-contained; throughout the text we
simply write $m = m_\pi$.
The reflection symmetry of the integrand about the vertical axis is guaranteed
for any physically viable form factor, and implies $\delta G_L^{A,-}(\tau) = \delta
G_L^{A,+}(\tau)$, as discussed in Appendix~\ref{app:minus-equals-plus}.
}
\label{fig:Wick-rotation}
\end{figure*}

The remaining task is to bring $\delta G_L^A(\tau)$ into a form that is
exponentially decaying in $\tau$. To do so, we Wick rotate by folding the
$\mathbb R+i \mu$ contour into the upper half of the complex plane as shown in
Fig.~\ref{fig:Wick-rotation}, so that it runs along the two sides of the
branch cut appearing on the positive imaginary axis. To explain this in detail
we first separate
\begin{equation}
\delta G_L^A(\tau) = \delta G_L^{A,+}(\tau) + \delta G_L^{A,-}(\tau) \,,
\label{eq:deltaGL-A-plus-minus}
\end{equation}
where $\pm$ indicates the contribution from $\mathbb R^{\pm} + i \mu$,
respectively. As shown in Appendix~\ref{app:minus-equals-plus}, the two
contributions are equal, i.e.~$\delta G_L^{A,-}(\tau) = \delta
G_L^{A,+}(\tau)$.

Focusing on the positive branch, we have
\begin{widetext}
\begin{equation}
\delta G_L^{A,+}(\tau) = \frac{1}{6} \sum_{\boldsymbol n \neq \boldsymbol 0}
\mathrm{Im} \!\left[ e^{i\theta} \int_{0}^{\infty} \frac{\mathrm{d}x}{2\pi}
\frac{ e^{\, i (x e^{i\theta} + i\mu)\vert \tau \vert - \vert \boldsymbol n
\vert L \sqrt{m^2 + (x e^{i\theta} + i\mu)^2/4}} }{ 4\pi (x e^{i\theta} + i\mu)
\vert \boldsymbol n \vert L } \left(4m^2 + (x e^{i\theta} + i\mu)^2 \right)
F\!\left[-(x e^{i\theta} + i\mu)^2\right]^2 \right] \,,
\label{eq:deltaGL-A-theta}
\end{equation}
\end{widetext}
where we have spelled out the contour parametrization $k_3 = x e^{i\theta} +
i\mu$.

We next note that the integrand is analytic for all $x \in [0, \infty)$, $\mu \in
(0, 2m)$, and $\theta \in [0, \pi/2)$. This analyticity is manifest for ratios
and exponentials of polynomials provided that no poles arise in the specified
parameter ranges, as one can readily check here. Less obvious is the role of
the square-root function appearing in the exponent. We define the branch cut of
this square-root along the negative real axis of its argument and note that this
translates to the cut running along $ x e^{i\theta} + i\mu \in i [2m,
\infty)$ which is outside the specified parameter ranges.

It only remains to argue for the analyticity of $F[-(x e^{i\theta} + i\mu)^2]$.
The key observation is that the form factor has a square-root branch point in
the timelike region, generating two Riemann sheets for elastic two-pion
scattering. Writing $s$ for the squared centre-of-mass energy, the branch cut
lies along $s \in [4m^2,\infty)$. Physical scattering probes the form factor on
the first (physical) sheet, just above the cut, i.e.\ at $s+i0^+$. The
spacelike region is reached by analytic continuation to negative $s$ without
crossing the cut.

Defining $F(-k_3^2)$ in the spacelike region, the timelike form factor on the
first sheet is given by $F((\sqrt{s}+i0^+)^2)$ for $\sqrt{s}>2m$, understood
via analytic continuation. For $x\in[0,\infty)$, $\mu\in(0,2m)$, and
$\theta\in[0,\pi/2)$, the function $F[-(x e^{i\theta}+i\mu)^2]$ is evaluated on
the first sheet below the cut. We can thus take the limit $\theta \to \pi/2$ to
reach
\begin{widetext}
\begin{equation}
\delta G_L^{A,+}(\tau) = \frac{1}{6} \sum_{\boldsymbol n \neq \boldsymbol 0}
\mathrm{Im} \left[ \int_{2m}^{\infty} \frac{\mathrm{d}x}{2\pi} \frac{ e^{-x
\vert \tau \vert - i \vert \boldsymbol n \vert L \sqrt{x^2/4-m^2}} }{ 4\pi x
\vert \boldsymbol n \vert L } \left(4m^2 - x^2 \right) F((x- i 0^+)^2)^2
\right] \,,
\label{eq:deltaGL-A-plus}
\end{equation}
where we have shifted the integration variable so that $x$ now runs over $[2m,
\infty)$.
\end{widetext}

We now recall that physical scattering gives access to the form factor on the
first Riemann sheet, infinitesimally above the branch cut. In
Eq.~\eqref{eq:deltaGL-A-plus}, by contrast, $F$ is sampled infinitesimally
below the cut. The two locations are related via the Schwarz reflection
principle, which implies that $F((x - i 0^+)^2) = F((x + i 0^+)^2)^*$ for real
$x$. We next make use of Watson's theorem, which relates the phase of the form
factor to elastic pion-pion scattering. Defining $p = \sqrt{s/4 - m^2}$ and
adopting the convention that $s$ is shorthand for $(\sqrt{s} + i 0^+)^2$
whenever $s > 4m^2$ and unless otherwise specified, we have
\begin{equation}
F(s)^2 = \left | F (s) \right |^2 e^{2 i \delta_{\pi\pi}(p)} \,,
\end{equation}
where, in the elastic regime, $\delta_{\pi\pi}(p)$ is the pion-pion scattering
phase shift in the isospin-1 channel. For the case of
Eq.~\eqref{eq:deltaGL-A-plus}, we have
\begin{equation}
F((x - i 0^+)^2)^2 = \left | F (x^2) \right |^2 e^{-2 i \delta_{\pi\pi}(p(x))}
\bigg \vert_{p(x) = \sqrt{x^2/4 - m^2}}\,.
\end{equation}

Including the factor of two for $\delta G_L^{A,-}(\tau) = \delta
G_L^{A,+}(\tau)$ and performing the change of variables from $x$ to $p$ on the
full integral, we reach
\begin{widetext}
\begin{equation}
\delta G_L(\tau) = -\frac{1}{6} \sum_{\boldsymbol n \neq \boldsymbol 0}
\frac{1}{ \pi^2 \vert \boldsymbol n \vert L} \mathrm{Im} \int_{0}^{
\infty}{{\rm d}p \ \frac{ p^3 e^{- 2 \sqrt{p^2 + m^2}|\tau|}}{p^2 + m^2}}
\left | F \big (4(p^2 + m^2) \big ) \right |^2 e^{-2 i \delta_{\pi\pi}(p)}
e^{-i \vert \boldsymbol n \vert L p} + \varepsilon_L(\tau)\,,
\label{eq:deltaGL-final}
\end{equation}
where $\varepsilon_L(\tau) = \delta G_L^B(\tau) + \delta
G_L^{\mathrm{reg}}(\tau)$ is suppressed by the interactions and by $1/\sqrt{L}$ (see Appendix~\ref{app:B-term-subdominant}). As we will see in the following section, the point
is not that this second contribution can be neglected in general, but
rather that, for pion masses, interactions, and smearing kernels for which this
is justified, the Euclidean-correlator approach of
Refs.~\cite{Hansen:2019rbh,Hansen:2020whp} reduces to the same leading
finite-volume correction as the Lellouch-L\"uscher-Meyer representation.

\begin{figure*}
\centering
\includegraphics[width=0.87\textwidth]{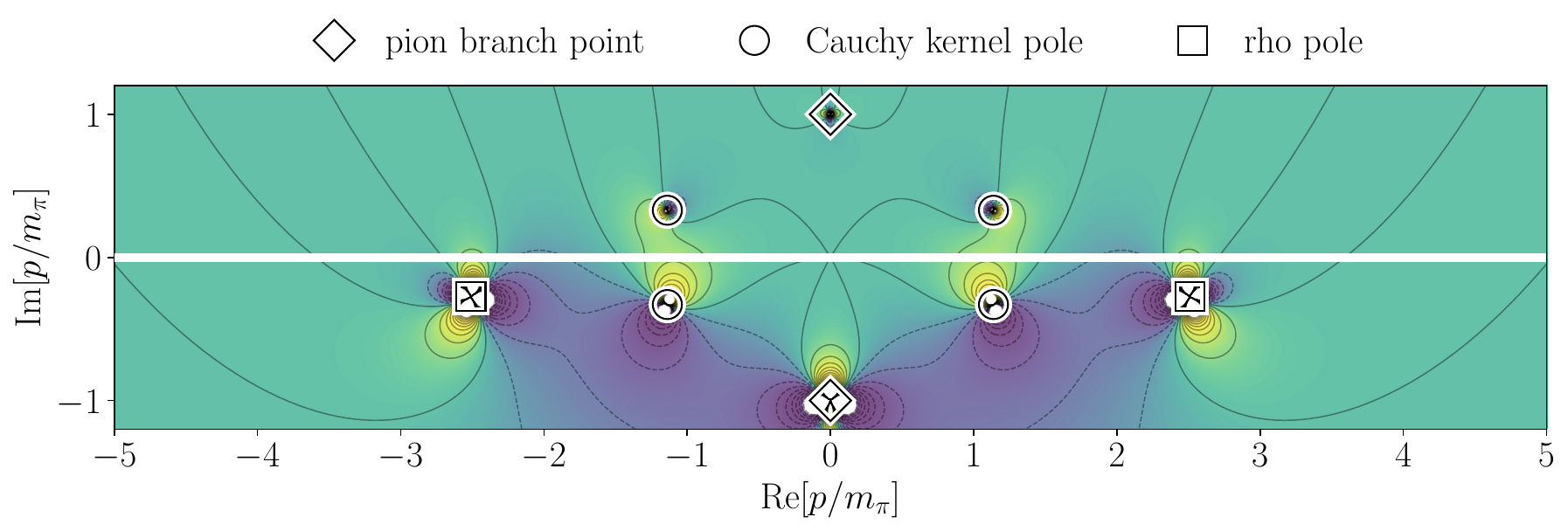}
\caption{Plot of the integrand appearing in Eq.~\eqref{eq:final-result} for
$\vert \boldsymbol n \vert L = 2/m$, for the Gounaris-Sakurai form factor (with
details and parameter choices given in Sec.~\ref{sec:numerics}) and the Cauchy
smearing kernel (Eq.~\eqref{eq:cauchy_kernel} with $\omega^*=3m$ and
$\sigma=0.5m$).}
\label{fig:p-form-factor}
\end{figure*}

Finally we can relate this to the smeared spectral density using
Eq.~\eqref{eq:kappa-expansion},
\begin{equation}
4 (p^2 + m^2) \sum_{i=1}^{N} c_{i}(\boldsymbol{\alpha}) \, e^{-2 \sqrt{p^2 +
m^2} \tau_i} = \widehat{\kappa}(2 \sqrt{p^2 + m^2}, \boldsymbol \alpha) \,.
\end{equation}
We deduce $\delta \widehat{\rho}_{L, \kappa}(\boldsymbol{\alpha}) =
f_{\kappa}(L,\boldsymbol{\alpha}) + \epsilon_{L, \kappa}(\boldsymbol{\alpha})$,
where the second term corresponds to the neglected terms, now in the context of
the smeared spectral density,
\begin{equation}
\epsilon_{L, \kappa}(\boldsymbol{\alpha}) = \sum_{i=
{1}}^{N} c_{i}(\boldsymbol{\alpha}) \, \varepsilon_L(\tau_i) \,,
\end{equation}
while the first term encodes the main result of this work:
\begin{equation}
f_{\kappa}(L,\boldsymbol{\alpha}) = \frac{1}{6} \sum_{\boldsymbol n \neq
\boldsymbol 0} \frac{1}{ 4 \pi^2 \vert \boldsymbol n \vert L} \text{Im}
\int_{0}^{ \infty}{{\rm d}p \frac{ p^3}{(p^2 + m^2)^2}} \left | F \big (4(p^2 +
m^2) \big ) \right |^2 \, e^{i 2 \delta_{\pi\pi}(p) + i \vert \boldsymbol n
\vert L p} \, \widehat{\kappa}(2 \sqrt{p^2+m^2}, \boldsymbol \alpha) \,.
\label{eq:final-result}
\end{equation}
\end{widetext}
This version assumes that $\widehat{\kappa}$ is real-valued. In the case of a
complex-valued $\widehat{\kappa}$, one should act the imaginary part only on
the exponential factor, rather than on the full integrand. Alternatively, one
can split the complex $\widehat{\kappa}$ into its real and imaginary parts and
write the result as a sum of two integrals, each of which has finite-volume
effects given by Eq.~\eqref{eq:final-result}. Note that we have kept the full
sum over $\boldsymbol n$, even though all but the first few terms scale
with an exponential in $L$ that is beyond the order we control.

To summarize, we have reached Eq.~\eqref{eq:final-result} by Wick rotating the
dominant part of the finite-volume correction derived in
Refs.~\cite{Hansen:2019rbh,Hansen:2020whp} into a form that is decaying in
$\tau$ and oscillatory in $L$, thereby allowing us to apply the coefficients
$c_i(\boldsymbol{\alpha})$ to the finite-volume correction and express the
latter in terms of $\widehat{\kappa}$.
This rotation also yields a dependence on the timelike
form factor. This is quite natural since, for a narrow smearing kernel, the
smeared spectral density is dominated by contributions from the timelike
region, a connection we make more precise in the following section.

A natural question, to be considered on a case-by-case basis, is whether
Eq.~\eqref{eq:final-result} can be rotated back to a representation that is
decaying in $L$ and which returns to the form factor in the spacelike region.
This will generally depend on the choice of smearing kernel and we leave a detailed investigation to future work. In Fig.~\ref{fig:p-form-factor}, we plot the integrand of Eq.~\eqref{eq:final-result} for a realistic model of the form factor (the Gounaris-Sakurai model, see Sec.~\ref{sec:numerics}) and for the Cauchy kernel defined in Eq.~\eqref{eq:cauchy_kernel} below. This $p$-space representation of the integrand brings both Riemann sheets of the form factor into view, and one can identify three categories of singularities. First, the points $p = \pm i m$, labelled ``pion branch point'' in the figure, at which the kinematic factor $1/(p^2+m^2)^2$ has a double pole and, in addition, the energy dependence of the Cauchy kernel, entering through $\omega = 2\sqrt{p^2+m^2}$, generates a square-root branch point. Second, poles corresponding to the $\rho$ resonance on the second sheet (i.e.~for $p$ with negative imaginary part), appearing in mirror pairs at $\pm \mathrm{Re}\,p$. Third, the poles of the smearing kernel, which for the Cauchy kernel lie at $p = \pm \sqrt{(\omega^* \pm i \sigma)^2/4 - m^2}$. A consequence of the Poisson summation in Eq.~\eqref{eq:final-result} is that the integration contour can only be closed in the upper half-plane, so that only the singularities at $p = im$ and the smearing-kernel poles with positive imaginary part directly contribute. The fact that the $\rho$-resonance poles are not directly encircled also has an interesting relation to causality arguments. In this context the observation is that, if the Fourier transform to a time-dependent correlator could be represented as a contour integral encircling a resonance pole, then the resulting correlator would exhibit unphysical time dependence. This in turn gives an argument as to why resonance poles cannot appear on the first Riemann sheet, i.e.~must have a negative imaginary part.

An illustrative alternative form of our main result follows by first defining
the two-pion contribution to the spectral density as
\begin{equation}
\rho_{\pi \pi}(\omega) = \frac{p^3}{6 \pi ^2 \omega^3}
\left | F ( \omega^2) \right |^2
\bigg \vert_{p = \sqrt{\omega^2/4 - m^2}} \,.
\end{equation}
Then we can write Eq.~\eqref{eq:final-result} as
\begin{equation}
f_{\kappa}(L,\boldsymbol{\alpha})
= \int {\rm d}\omega \, \rho_{\pi \pi}(\omega) \, \widehat{\kappa}(\omega,
\boldsymbol \alpha) \, \Delta(\omega, L) \,, \label{eq:final-result-2}
\end{equation}
where we have introduced
\begin{equation}
\Delta(\omega, L) = \sum_{\boldsymbol n \neq \boldsymbol 0} \frac{ \sin(2
\delta_{\pi\pi}(p) + \vert \boldsymbol n \vert L p) }{\vert \boldsymbol n \vert
L p} \bigg \vert_{ p = \sqrt{\omega^2/4-m^2}} \,.
\end{equation}

This expression illustrates that the finite-volume correction can be expressed
as an integral over the infinite-volume spectral density, weighted by the
smearing kernel and a function $\Delta(\omega, L)$ that encodes the
finite-volume effects. Future work is needed to determine whether this form is
particularly useful, e.g.~as a means of spectrally reconstructing the
finite-volume correction itself or as a step towards extending these results to
more complicated systems, such as those involving three or more particles.

We close by considering whether our main result applies for non-linear
reconstruction methods of $\widehat{\rho}_{L, \kappa}$. At face value, the
derivation relies on the linearity of the relation between $\widehat{\rho}_{L,
\kappa}$ and $G_L(\tau)$, as expressed in Eq.~\eqref{eq:G-to-rho-hat}. However,
as the smeared spectral density is, in and of itself, a perfectly well-defined
quantity, the expression for its finite-volume effects should be independent of
the method used to reconstruct it. Therefore, the main result must hold
for non-linear reconstruction methods as well. However, imperfections in the estimator
for $\widehat{\rho}_{L, \kappa}$ could propagate in highly non-trivial ways
into the finite-volume correction. For linear reconstruction, by contrast, one
can replace $\widehat \kappa$ by $\widehat{\kappa}_{\sf r}$ in $\delta \widehat
\rho_{L,\kappa}$, if the two differ enough to significantly affect the value of
the latter.

\section{Relation to the approach of Lellouch-L\"uscher and
Meyer}\label{sec:LL}

An alternative approach to quantify finite-volume effects on Euclidean
correlators, which readily generalizes to spectral densities, is provided by
the formalism of Lellouch and L\"uscher~\cite{Lellouch:2000pv} and the
work of Meyer~\cite{Meyer:2011um}. In this section, we relate our result to
that approach.

The starting point is to insert a complete set of finite-volume energy
eigenstates between the current operators in Eq.~\eqref{eq:GL-vv-def}, yielding
\begin{equation}
G_L(\tau) = \frac{1}{3} \sum_{n} c_{n}(L) \, e^{- E_n(L) \tau} \,,
\end{equation}
where we have introduced
\begin{equation}
c_{n}(L) = L^3 \sum_{k=1}^{3}
\vert \langle n, \boldsymbol 0 \vert j_k(0) \vert0 \rangle_L \vert^2 \,,
\end{equation}
and where the overall minus sign in the original definition is absorbed, using
the fact that the Euclidean vector current is anti-Hermitian.
Here $E_n(L)$ and $\vert n, \boldsymbol 0 \rangle_L$ are the finite-volume
energy eigenvalues and states with zero total momentum, respectively. The
states are normalized to unity.

An advantage of this representation is that one can immediately write down the
corresponding smeared spectral density,
\begin{equation}
\widehat{\rho}_{L, \kappa}(\boldsymbol{\alpha}) = \frac{1}{3} \sum_{n=1}^\infty
\frac{c_{n}(L)}{E_n(L)^2} \, \widehat{\kappa}(E_n(L), \boldsymbol \alpha) \,.
\label{eq:rho-hat-LL}
\end{equation}

To make progress we now use L\"uscher's quantization
condition~\cite{Luscher:1986pf,Luscher:1990ux} which relates the energies
$E_n(L)$ to infinite-volume scattering amplitudes.
More precisely, the finite-volume states that appear here transform as an
irreducible representation, denoted by $T_1^-$, of the 48-element octahedral
group (the appropriate symmetry group for states with zero total momentum in a
cubic finite volume). In the context of the continuous rotations of the
infinite-volume theory, this representation contains the $p$-wave ($\ell$=1)
scattering channel as well as an infinite tower of higher partial waves.
Formally all of these contribute to the finite-volume energies. But in
practice, the $p$-wave channel dominates the low-energy scattering of two pions
with isospin 1.

In the case that only this dominant scattering channel is included, the
finite-volume energies are related to the corresponding scattering phase shift
$\delta_{\pi\pi}(p(E))$ according to the following quantization condition:
\begin{equation}
\mathcal Q(E_n,L) = \pi n \,,
\label{eq:Q-based-QC}
\end{equation}
where
\begin{align}
\mathcal Q(E,L) & = \delta_{\pi\pi}(p(E)) + \phi(q(E,L)) \,,
\label{eq:Q}
\\
\cot \phi(q) & = - \frac{\mathcal Z_{00}(1,q^2)}{q \pi^{3/2}} \,,
\end{align}
and we have introduced the kinematic variables
\begin{equation}
p(E) = \sqrt{E^2/4-m^2} \,, \qquad q(E,L)= \frac{p(E)L}{2\pi} \,.
\label{eq:q-def}
\end{equation}
The function $\phi(q)$
is also defined to be a continuous function of $q$, i.e.~we choose the branch
of the inverse cotangent accordingly. See Fig.~\ref{fig:phi-q-vs-q-squared} in App.~\ref{app:phi_expansion_and_positivity}.

Writing the quantization condition in the form of Eq.~\eqref{eq:Q-based-QC}
gives an unambiguous meaning to the label $n$ of the finite-volume energy
levels. As indicated by the sum in Eq.~\eqref{eq:rho-hat-LL}, $n$ in this
convention, and for this particular system, runs over all positive integers.
This is further clarified in App.~\ref{app:phi_expansion_and_positivity}.

The L\"uscher zeta function, $\mathcal Z_{00}(1,q^2)$ can be defined in a
theoretically useful (though not numerically efficient) way via the regulated
sum
\begin{equation}
\sqrt{4\pi} \mathcal Z_{00}(1,q^2) = \lim_{\Lambda \to \infty}
\bigg[ \sum_{\boldsymbol n}^{\vert \boldsymbol n \vert < \Lambda} \frac{1}{
\boldsymbol n^2-q^2} - 4\pi \Lambda \bigg] \,,
\label{eq:z00def}
\end{equation}
where the sum runs over all triplets of integers $\boldsymbol n \in \mathbb
Z^3$.

In addition to providing a relation between the scattering phase shift and the
finite-volume energies, the same function $\mathcal Q(E,L)$ also relates
finite- and infinite-volume matrix
elements~\cite{Lellouch:2000pv,Meyer:2011um,Briceno:2014uqa,Briceno:2015csa}.
Taking into account the relation
between the infinite-volume matrix elements and the timelike pion form factor
$F(E^2)$, one finds
\begin{equation}
c_n(L) = 4 p_n^2 \vert F(E_n^2) \vert^2 \frac{p_n}{8 \pi E_n }
\frac{1}{\mathcal Q^{(1,0)}(E_n,L)} \,,
\end{equation}
where the superscript in $\mathcal Q^{(1,0)}$ indicates a partial derivative
with respect to the first argument.

Substituting these results into Eq.~\eqref{eq:rho-hat-LL}, we reach
\begin{align}
\begin{split}
\widehat{\rho}_{L, \kappa}(\boldsymbol{\alpha}) & = \frac{1}{3}
\sum_{n=1}^\infty \int_{2m}^\infty \! \! d E \, \frac{\delta \big (E - E_n(L)
\big )}{\mathcal Q^{(1,0)}(E,L)} \\[-5pt] & \hspace{60pt} \times
\frac{p(E)^3}{2 \pi E^3} \, \vert F(E^2) \vert^2 \, \widehat{\kappa}(E,
\boldsymbol{\alpha}) \,,
\end{split}
\label{eq:rho-hat-LL-2}
\\
\begin{split}
& = \frac{1}{3} \sum_{n=-\infty}^\infty \int_{2m}^\infty \! \! d E \, \delta
\big (n \pi - \mathcal Q(E, L) \big ) \\[-5pt] & \hspace{60pt} \times
\frac{p(E)^3}{2 \pi E^3} \, \vert F(E^2) \vert^2 \, \widehat{\kappa}(E,
\boldsymbol{\alpha}) \,.
\end{split}
\label{eq:rho-hat-LL-3}
\end{align}
Note that $E$ is equivalent to $\sqrt s$ introduced earlier.
In the second step we have used the property of the Dirac delta function under
a change of variables to rewrite the first factor of the integrand. In addition
we have extended the sum over $n$ to all integers, noting that the
non-positive-$n$ terms do not contribute since $\mathcal Q(E,L) > 0$ for all $E
> 2m$. This is explained in
App.~\ref{app:phi_expansion_and_positivity}.\footnote{A subtlety arises from
the fact that $\mathcal Q(2m, L) = 0$ for all $L$. However, the contribution from this
point is removed since the integrand vanishes at $E=2m$ due to the $p^3$
factor. This vanishing is not a genuine finite-volume energy but rather an
artifact of the $p$-wave phase space in the definition of $\mathcal Q(E,L)$.}

Equation~\eqref{eq:rho-hat-LL-3} closely follows Appendix C of
Ref.~\cite{Bulava:2021fre}. The utility of these manipulations is that we can
now apply the Poisson summation formula
\begin{equation}
\sum_{n=-\infty}^\infty \delta \big (n \pi - \mathcal Q(E, L) \big ) \\ =
\frac{1}{\pi} \sum_{l = - \infty}^\infty e^{2 i l \mathcal Q(E,L) } \,.
\end{equation}

We find that the finite-volume correction to the smeared spectral density can
be written as an infinite sum over Poisson dual modes, where the subtraction of
the infinite-volume result removes the $l=0$ term:
\begin{equation}
\delta \widehat{\rho}_{L, \kappa}(\boldsymbol{\alpha}) = 2 \text{Re} \sum_{l =
1}^\infty \widehat{\rho}^{(l)}_{L, \kappa}(\boldsymbol{\alpha}) \,,
\label{eq:rhok}
\end{equation}
and an individual term is given by
\begin{multline}
\widehat{\rho}^{(l)}_{L, \kappa}(\boldsymbol{\alpha}) = \frac{1}{3} \frac{1}{16
\pi^2} \int_{-\infty}^\infty \! d p \, e^{2 i l \mathcal Q(E(p),L)} \,
\frac{p^4}{ (p^2 + m^2)^2} \\ \times \vert F(E(p)^2) \vert^2 \,
\widehat{\kappa}(E(p), \boldsymbol{\alpha}) \,.
\label{eq:rhokDef}
\end{multline}
Here we have additionally performed the change of integration variables from
$E$ to $p$ and extended the integration range to the full real line by noting
that the real part of the integrand is an even function of $p$.

While $\mathcal Q(E(p),L)$ has an ill-defined $L \to \infty$ limit for
real-valued $p$, this is resolved by deforming the integration contour into the
complex plane. In particular, if we assume $\text{Im}(p) = \mu > 0$, we can use
the exponentially convergent representation of the zeta function given in
Refs.~\cite{Davoudi:2011md,Hansen:2016ync,Briceno:2019nns}. As detailed in
App.~\ref{app:phi_expansion_and_positivity}, this leads to the large-$L$
expansion
\begin{equation}
e^{2 i \mathcal Q(E(p),L)} = \frac{ e^{2i\delta_{\pi\pi}(p)}}{2 ipL}
\sum_{\boldsymbol n \neq \boldsymbol 0} \frac{e^{i \vert \boldsymbol n \vert
pL} }{\vert \boldsymbol n \vert} + \mathcal O(e^{{2}ipL }/(Lp)^2) \,.
\label{eq:QC_expansion}
\end{equation}
Substituting this result into Eq.~\eqref{eq:rhokDef}, we find that the leading
behaviour at large $L$ is captured by
\begin{widetext}
\begin{equation}
\widehat{\rho}^{(1)}_{L, \kappa}(\boldsymbol{\alpha}) = \frac{1}{3} \frac{1}{16
\pi^2} \int_{\mathbb R + i \mu} \! d p \, \bigg [
\frac{ e^{2i\delta_{\pi\pi}(p)}}{2 ipL} \sum_{\boldsymbol n \neq \boldsymbol 0}
\frac{e^{i \vert \boldsymbol n \vert pL} }{\vert \boldsymbol n \vert} +
\mathcal O(e^{{2}ipL }/(Lp)^2)
\bigg ] \, \frac{p^4}{ (p^2 + m^2)^2} \, \left | F \big (4(p^2 + m^2) \big )
\right |^2 \, \widehat{\kappa}(E(p), \boldsymbol{\alpha}) \,.
\label{eq:rho1v1}
\end{equation}
\end{widetext}
It follows that the leading finite-volume correction predicted by the
Lellouch-L\"uscher-Meyer approach exactly matches the leading term in our main
result of the previous section, Eq.~\eqref{eq:final-result}:
\begin{align}
2 \text{Re} \widehat{\rho}^{(1)}_{L, \kappa}(\boldsymbol{\alpha})
& =
f_{\kappa}(L,\boldsymbol{\alpha})
+ \epsilon^{(1)}_{L, \kappa}(\boldsymbol{\alpha})
\label{eq:rho1v3} \,,
\end{align}
where $\epsilon^{(1)}_{L, \kappa}(\boldsymbol{\alpha})$ collects the subleading
corrections arising from the $\mathcal O [e^{2i L p}/(Lp)^2 ]$ term in
Eq.~\eqref{eq:rho1v1}. A subtlety here is that, after separating the leading
and subleading contributions, we have deformed the integration contour back to
the real axis. This is valid since the value of this integral is unchanged by
the contour deformation. Thus Eq.~\eqref{eq:rho1v3} is expressed in terms of the same universal function $f_{\kappa}(L,\boldsymbol{\alpha})$ as the main result of the previous section, defined in Eq.~\eqref{eq:final-result}.

Interestingly the contribution identified in this section comes entirely from
the $l=1$ Poisson dual mode. The $l>1$ modes contribute higher powers of the
$S$-matrix ($e^{2 i l \delta_{\pi\pi}(p)}$) and it would be interesting to
understand whether these can be related to the subleading corrections in the
main result. We leave this question to future work.

\section{Numerical results}\label{sec:numerics}

\begin{figure*}
\centering
\includegraphics[width=0.9\textwidth]{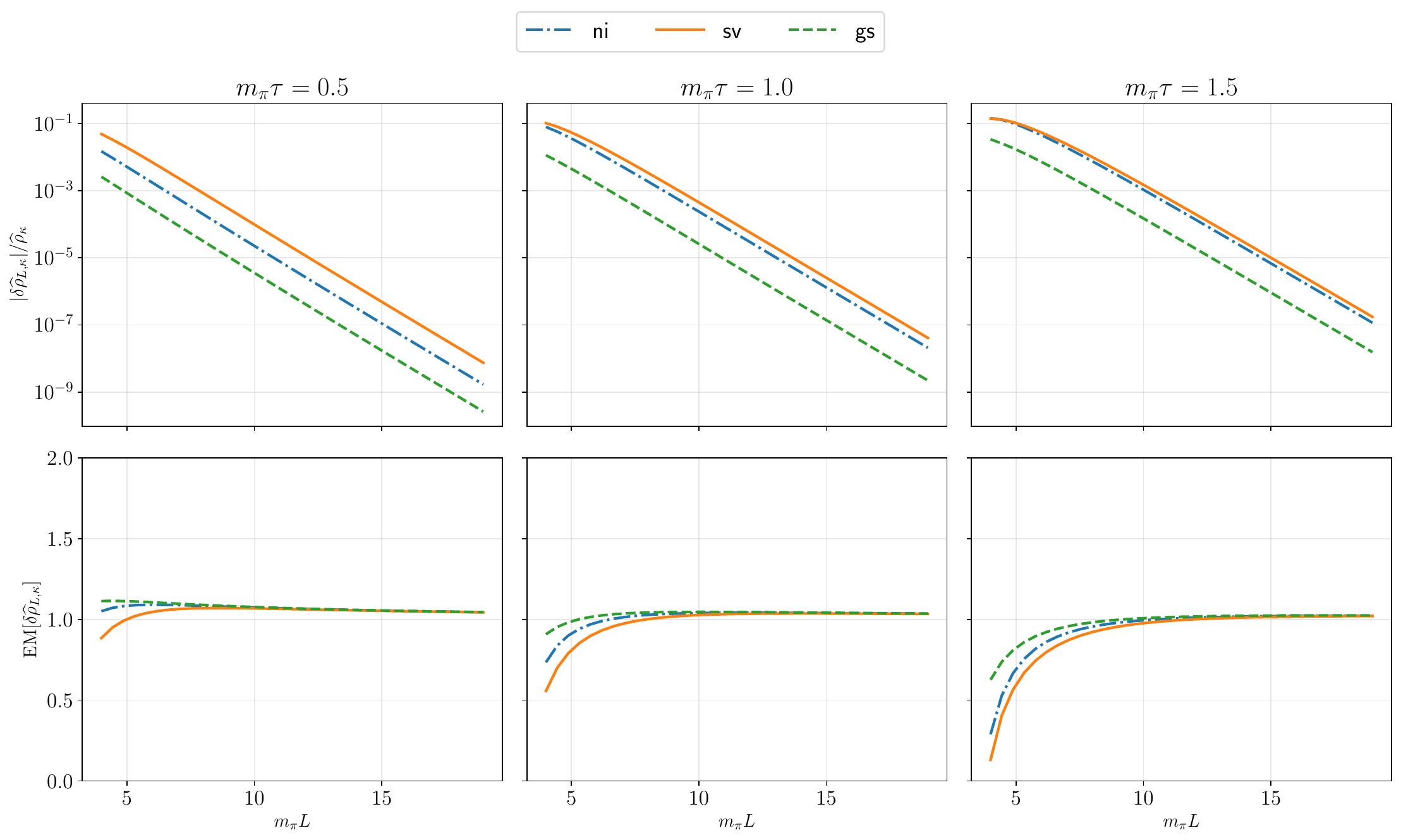}
\caption{
Finite-volume effects on the Euclidean vector-vector correlator.
The top row shows the relative finite-volume corrections at three Euclidean
time separations, as indicated.
The bottom row presents the same effects expressed as an effective mass,
defined in Eq.~\eqref{eq:eff_mass} of the main text.
In each panel, results are shown for three models of the timelike pion form
factor: non-interacting (\textsf{ni}), scattering volume (\textsf{sv}), and
Gounaris-Sakurai (\textsf{gs}), as indicated in the legend. These models are defined in the main text in Eqs.~\eqref{eq:ni_case}, ~\eqref{eq:sv_case}, and ~\eqref{eq:gs_case}, respectively.
}
\label{fig:laplacian}
\end{figure*}

\begin{table*}
\begin{center}
\addtolength{\leftskip}{-2cm}
\addtolength{\rightskip}{-2cm}
\setlength{\tabcolsep}{2pt}
$ |\delta \widehat{\rho}_{L, \kappa}|/\widehat{\rho}_\kappa$ \\[8pt]
\begin{tabular}{l|c||c|c|c|c|c|c|c|c||c}
& $mL$ &
$|\boldsymbol n|=1$ & $\sqrt{2}$ & $\sqrt{3}$ &
$2$ & $\sqrt{5}$ & $\sqrt{6}$ & $2\sqrt{2}$ & $3$ &
$\sum_{\boldsymbol n}$ \\
\hline\hline

& 4 & 0.147 & 0.119 & 0.0270 & 0.00749 & 0.0121 & 0.00521 & $5.74\times10^{-4}$
& $7.18\times10^{-4}$ & 0.318 \\
& 5 & 0.0950 & 0.0355 & 0.00530 & 0.00107 & 0.00131 & $4.48\times10^{-4}$ &
$3.27\times10^{-5}$ & $3.41\times10^{-5}$ & 0.139 \\
$\sf{ni}$ \quad & 6 & 0.0451 & 0.00941 & $9.63\times10^{-4}$ &
$1.44\times10^{-4}$ & $1.37\times10^{-4}$ & $3.71\times10^{-5}$ &
$1.82\times10^{-6}$ & $1.59\times10^{-6}$ & 0.0558 \\
& 7 & 0.0190 & 0.00236 & $1.69\times10^{-4}$ & $1.89\times10^{-5}$ &
$1.40\times10^{-5}$ & $3.04\times10^{-6}$ & $1.01\times10^{-7}$ &
$7.37\times10^{-8}$ & 0.0215 \\
& 8 & 0.00749 & $5.74\times10^{-4}$ & $2.91\times10^{-5}$ &
$2.46\times10^{-6}$ & $1.42\times10^{-6}$ & $2.48\times10^{-7}$ &
$5.58\times10^{-9}$ & $3.42\times10^{-9}$ & 0.00809 \\
\hline\hline
& 4 & 0.140 & 0.138 & 0.0342 & 0.00994 & 0.0165 & 0.00727 & $8.21\times10^{-4}$
& $1.04\times10^{-3}$ & 0.348 \\
& 5 & 0.104 & 0.0454 & 0.00719 & 0.00149 & 0.00187 & $6.47\times10^{-4}$ &
$4.80\times10^{-5}$ & $5.03\times10^{-5}$ & 0.161 \\
$\sf{sv}$ \quad & 6 & 0.0539 & 0.0127 & 0.00136 & $2.07\times10^{-4}$ &
$2.00\times10^{-4}$ & $5.47\times10^{-5}$ & $2.71\times10^{-6}$ &
$2.37\times10^{-6}$ & 0.0685 \\
& 7 & 0.0241 & 0.00330 & $2.44\times10^{-4}$ & $2.77\times10^{-5}$ &
$2.07\times10^{-5}$ & $4.53\times10^{-6}$ & $1.51\times10^{-7}$ &
$1.11\times10^{-7}$ & 0.0277 \\
& 8 & 0.00994 & $8.21\times10^{-4}$ & $4.27\times10^{-5}$ &
$3.64\times10^{-6}$ & $2.12\times10^{-6}$ & $3.72\times10^{-7}$ &
$8.40\times10^{-9}$ & $5.16\times10^{-9}$ & 0.0108 \\
\hline\hline
& 4 & 0.0338 & 0.0195 & 0.00407 & 0.00109 & 0.00171 & $7.29\times10^{-4}$ &
$7.88\times10^{-5}$ & $9.81\times10^{-5}$ & 0.0611 \\
& 5 & 0.0169 & 0.00533 & $7.56\times10^{-4}$ & $1.49\times10^{-4}$ &
$1.81\times10^{-4}$ & $6.10\times10^{-5}$ & $4.41\times10^{-6}$ &
$4.58\times10^{-6}$ & 0.0234 \\
$\sf{gs}$ \quad & 6 & 0.00720 & 0.00135 & $1.34\times10^{-4}$ &
$1.96\times10^{-5}$ & $1.85\times10^{-5}$ & $5.00\times10^{-6}$ &
$2.43\times10^{-7}$ & $2.12\times10^{-7}$ & 0.00872 \\
& 7 & 0.00285 & $3.29\times10^{-4}$ & $2.31\times10^{-5}$ & $2.55\times10^{-6}$
& $1.88\times10^{-6}$ & $4.06\times10^{-7}$ & $1.34\times10^{-8}$ &
$9.77\times10^{-9}$ & 0.00321 \\
& 8 & 0.00109 & $7.88\times10^{-5}$ & $3.94\times10^{-6}$ & $3.29\times10^{-7}$
& $1.90\times10^{-7}$ & $3.30\times10^{-8}$ & $7.37\times10^{-10}$ &
$4.52\times10^{-10}$ & 0.00117 \\
\hline
\end{tabular}
\end{center}
\caption{Numerical estimates of the contributions from the first few single
shells to Eq.~\eqref{eq:final-result} for the Euclidean vector-vector
correlator at $m \tau=1.5$, shown for the three form-factor parametrizations
(\textsf{ni}), (\textsf{sv}), and (\textsf{gs}), defined in the main text. }

\label{tab_laplacian}
\end{table*}

\begin{figure*}
\centering
\includegraphics[width=0.9\textwidth]{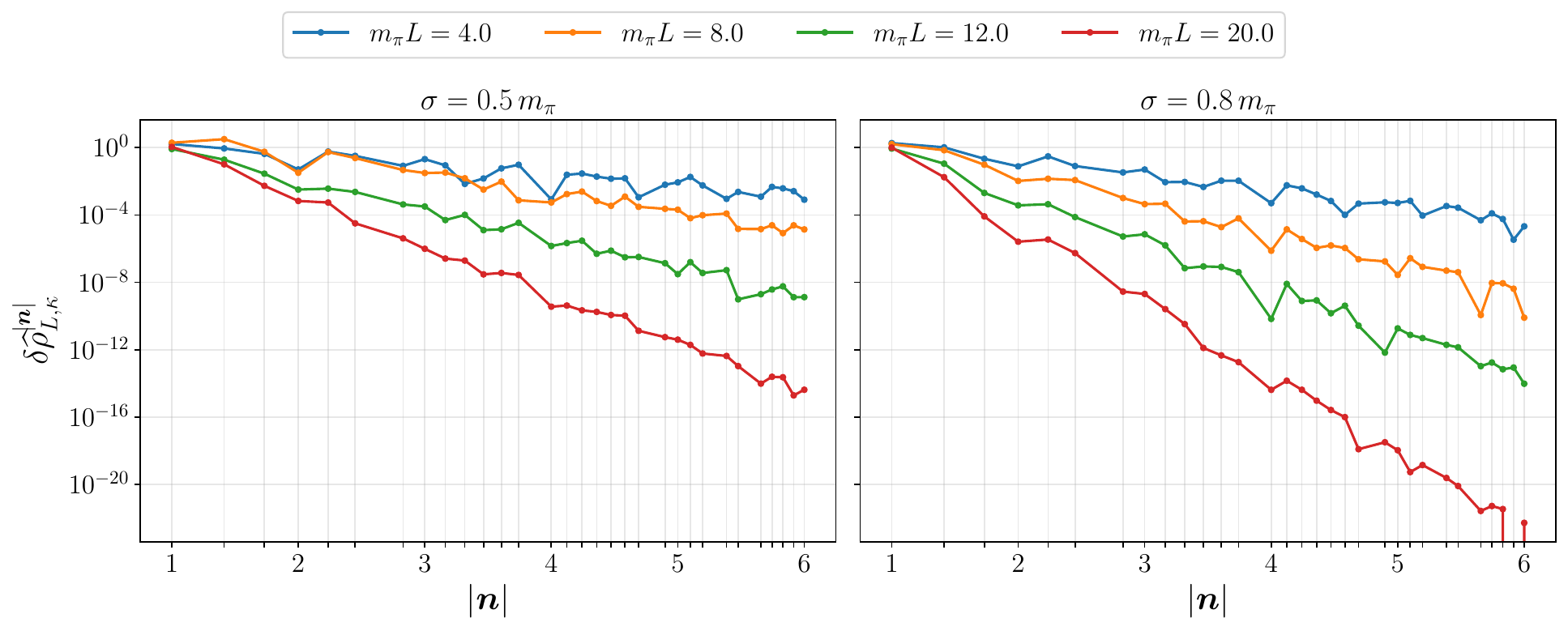}
\caption{
Fixed-$|\boldsymbol{n}|$ contributions to $f_{\kappa}(L, \boldsymbol{\alpha})$,
normalized with respect to their cumulative sum. Here we consider a Cauchy
smearing kernel, defined in Eq.~\eqref{eq:cauchy_kernel}, and centred at
$\omega^* = 3m$, together with the Gounaris-Sakurai form-factor parametrization
$(\sf{gs})$.}
\label{fig:convergence_cauchy}
\end{figure*}

\begin{figure*}
\centering
\includegraphics[width=0.7\textwidth]{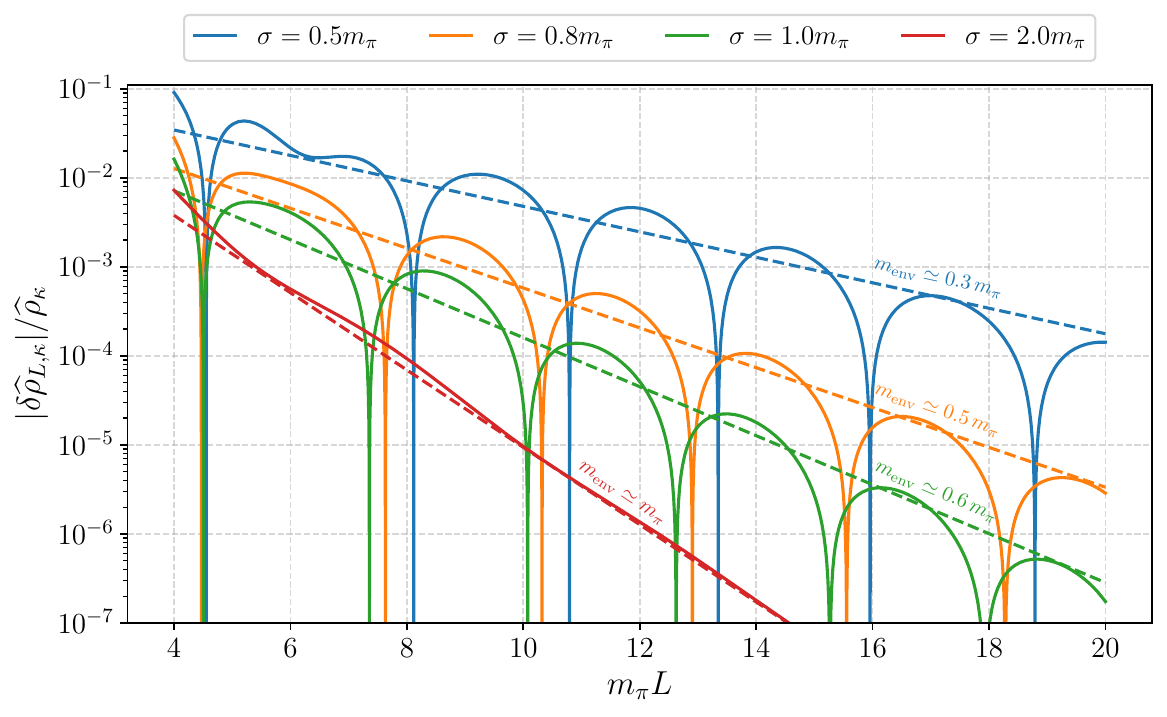}
\caption{
Cumulative finite-volume effects from shells with $|\boldsymbol{n}| \leq 6$ on
the Cauchy-smeared spectral density, evaluated using the Gounaris-Sakurai
form-factor parametrization $(\sf{gs})$. The Cauchy smearing kernel is centred
at $\omega^* = 3 m$, and results are shown for several values of the smearing
width $\sigma$.}
\label{fig:cauchy_fve}
\end{figure*}

We now turn to the numerical evaluation of our main result. We perform this
exercise for three different choices of the form factor, $F$, and for two
possible smearing kernels, $\widehat{\kappa}$.

We first introduce the three form-factor models.
Beginning with the noninteracting case, obtained by setting
\begin{equation}
 \vert F(4(p^2+m^2)) \vert^2 e^{2i\delta_{\pi\pi}(p)} = 1 \,,
\label{eq:ni_case}
\end{equation}
we write
\begin{multline}
\delta \widehat{\rho}^\mathrm{(\sf{ni})}_{L, \kappa}(\boldsymbol{\alpha}) =
\sum_{\boldsymbol n \neq \boldsymbol 0} \frac{1}{48 \pi^2 \vert \boldsymbol n
\vert L} \mathrm{Im} \int_{-\infty}^{ \infty}{{\rm d}p \frac{ p^{3}}{(p^2 +
m^2)^2}} \\ \times \, e^{i \vert \boldsymbol n \vert L p} \, \widehat{\kappa}(2
\sqrt{p^2+m^2}, \boldsymbol \alpha) \,,
\label{eq:final-result-ni}
\end{multline}
where $\sf{ni}$ indicates the noninteracting case and where we have extended
the integral to negative $p$, taking advantage of the evenness of the
integrand.

As a quick aside, we note that it is particularly straightforward to connect
this simplified quantity with the Lellouch-L\"uscher-Meyer perspective.
Although we have done this for general theories in the previous section, we
think it is instructive to see how it works here. We start by writing
\begin{equation}
\delta \widehat{\rho}^\mathrm{(\sf{ni})}_{L, \kappa}(\boldsymbol{\alpha}) =
\widehat{\rho}^\mathrm{(\sf{ni})}_{L, \kappa}(\boldsymbol{\alpha}) - \lim_{L
\to \infty} \widehat{\rho}^\mathrm{(\sf{ni})}_{L, \kappa}(\boldsymbol{\alpha})
\,,
\end{equation}
where
\begin{multline}
\widehat{\rho}^\mathrm{(\sf{ni})}_{L, \kappa}(\boldsymbol{\alpha}) = -
\frac{1}{3} \sum_{k=1}^3 \int {\rm d}^3 \boldsymbol x \\ \times \langle 0 \vert
j_k(0) e^{i \widehat {\boldsymbol P} \cdot \boldsymbol x} \widehat H^{-2}
\widehat{\kappa}(\widehat H, \boldsymbol \alpha) j_k(0) \vert 0 \rangle_L \,.
\end{multline}
We then insert a complete set of states $\sum_n \vert n \rangle_L
\langle n \vert_L$ between the current operators, where in the noninteracting
case the two-pion states are given by
\begin{multline}
\sum_n \vert n \rangle_L \langle n \vert_L = \frac{1}{L^6} \sum_{\boldsymbol p,
\boldsymbol p'} \frac{1}{2 \omega_{\boldsymbol p}} \frac{1}{2
\omega_{\boldsymbol p'}} \\ \times \vert \pi^+(\boldsymbol p) \pi^-(\boldsymbol
p') \rangle_L \langle \pi^+(\boldsymbol p) \pi^-(
\boldsymbol p') \vert_L \,,
\end{multline}
with $\omega_{\boldsymbol p} = \sqrt{\boldsymbol p^2 + m^2}$. Using
\begin{equation}
\langle 0 \vert j_k(0) \vert \pi^+(\boldsymbol p) \pi^-(\boldsymbol p')
\rangle_L = -i (\boldsymbol p - \boldsymbol p')_k \,,
\end{equation}
and applying the Poisson summation formula gives
\begin{equation}
\delta \widehat{\rho}^\mathrm{(\sf{ni})}_{L, \kappa}(\boldsymbol{\alpha}) =
\frac{4}{3} \sum_{\boldsymbol n \neq \boldsymbol 0} \int \! \frac{{\rm d}^3
\boldsymbol p}{(2 \pi)^3 (2 \omega_{\boldsymbol p})^4} e^{i \boldsymbol p \cdot
\boldsymbol n L}
\vert \boldsymbol{p} \vert^2
\widehat{\kappa}(2 \omega_{\boldsymbol p}, \boldsymbol \alpha) \,.
\end{equation}
Evaluating the angular integrals
then recovers Eq.~\eqref{eq:final-result-ni}.

We next introduce an interacting case, where the form factor is modelled
via the leading order term in a small momentum expansion of the K-matrix. The
starting point is an expression for the form factor in the two-particle elastic
regime based on unitarity and analyticity,
\begin{equation}
F(s) = \frac{1}{1 - i \rho(s) \mathcal K(s)} \mathcal A(s) \,,
\end{equation}
where $\rho(s) = p/(16 \pi \sqrt{s})$ (with $p = \sqrt{s/4 - m^2}$) is the
two-particle phase space factor, $\mathcal K(s)$ is the K-matrix, and $\mathcal
A(s)$ is an analytic function describing the coupling of the current to the
two-pion state.

For the discussion here it is instructive to be explicit about the routing of
the branch cut and the definition of the first and second Riemann sheets. To
this end we define
\begin{equation}
(z)^{1/2}_{\mathbb R^+}= |z|^{1/2} \exp\!\left(\tfrac{i}{2}\text{Arg}(z)
\right),
\quad
0 \leq \text{Arg}(z) < 2\pi \,,
\end{equation}
and note that this implies that $\text{Im}\! \big [ (s - 4 m^2)^{1/2}_{\mathbb
R^+} \big ] \geq 0$ for all $s \in \mathbb C$. This definition serves to define
$F(s)$ on the first Riemann sheet.

Setting $\mathcal K(s) = 16 \pi \sqrt{s}(a_1 p^2)$, with $a_1$ the scattering
volume, and taking $\mathcal A(s) = 1$ for simplicity, we have
\begin{equation}
F^{I}_{\sf{sv}}(s) = \frac{1}{1 - i a_1 [s/4-m^2]^{3/2}_{\mathbb R^+} } \,,
\label{eq:sv_case}
\end{equation}
where the superscript $I$ indicates the first Riemann sheet and the subscript
$\sf{sv}$ stands for ``scattering volume''.
The corresponding expression for the second Riemann sheet is obtained by
replacing $i$ with $-i$ in the denominator.
However, in the present context only the first expression is required.
This model is a simple generalization of the noninteracting case. While it is
not particularly accurate phenomenologically, it is a useful test case for our
formalism and allows us to explore the effect of interactions on the
finite-volume corrections. In particular, the scattering volume $a_1$ controls
the strength of the interaction, with $a_1 = 0$ corresponding to the
noninteracting case. We set $a_1 m^3 = 0.1$ for the numerical evaluations
presented below.

For our third and final example we consider the Gounaris-Sakurai (GS) form
factor model~\cite{Gounaris:1968mw}, which provides a phenomenologically
successful description of the pion electromagnetic form factor in the timelike
region dominated by the $\rho$ resonance as well as in the spacelike region, here taken without isospin breaking effects such as $\rho-\omega$ mixing.

The essential idea is to include a term in the inverse K-matrix of the form
$(s/4 - m^2) h(\sqrt{s})$, where $h(\sqrt{s})$ is the once-subtracted
Chew-Mandelstam function, defined as
\begin{equation}
h(\sqrt{s}) = \frac{2}{\pi}\,
\frac{p}{\sqrt{s}}\,
\log\!\left(
\frac{\sqrt{s} + 2 p}{2 m}
\right) \bigg \vert_{p = (s/4 - m^2)^{1/2}_{\mathbb R^+}} \,,
\end{equation}
and is motivated by a dispersion relation for the real part of the two-pion
loop. The full form factor is then constructed to ensure the correct
normalization at $s=0$ as well as the presence of the $\rho$ resonance at $s =
m_\rho^2$. The final expression is
\begin{widetext}
\begin{align}
\label{eq:gs_case}
F_{\sf{gs}}^{I}(s)
&=
\frac{f_0}{
\left({s}/{4}-m^2\right) h\!\left(\sqrt{s}\right)
- (m_\rho^2/4 - m^2) h(m_\rho)
+ b \left(
{s}/{4} - m_\rho^2/4
\right)
-
i {
[s/4-m^2]^{3/2}_{\mathbb R^+}
}/{
\sqrt{s}}
},
\\[6pt]
f_0
&=
-\frac{m^{2}}{\pi}
- (m_\rho^2/4 - m^2) h(m_\rho)
- b\,\frac{m_\rho^{2}}{4}, \notag
\\[6pt]
b
&=
-h(m_\rho)
- \frac{24 \pi}{g_{\rho\pi\pi}^{2}}
- \frac{2 (m_\rho^2/4 - m^2)}{m_\rho}\, h'(m_\rho) \,, \notag
\end{align}
\end{widetext}
where the subscript $\sf{gs}$ stands for Gounaris-Sakurai and the superscript
$I$ indicates the first Riemann sheet. The parameters of the model are the
$\rho$ mass, $m_\rho$, and the $\rho \pi \pi$ coupling, $g_{\rho\pi\pi}$, which
controls the width of the resonance. We set these parameters approximately to
their physical values, $m_\rho/m = 5.0$ and $g_{\rho\pi\pi} =
6.0$~\cite{ParticleDataGroup:2024cfk}.

In Fig.~\ref{fig:laplacian} we show the relative finite-volume corrections of
the Euclidean correlator, $\delta G_L(\tau)/G(\tau)$ (equivalently on the
spectral density smeared with the exponential kernel $\widehat{\kappa}(\omega,
\boldsymbol \alpha) = \omega^2 e^{- \omega \tau}$).
The results are shown as functions of $L$ for three different values of $\tau$
and for the three models of the form factor described above.
To illustrate the energy scale that dominates the asymptotic behaviour, we
calculate the corresponding effective mass, defined as the logarithmic
derivative of $\delta \widehat{\rho}_{L, \kappa}$,
\begin{equation}
 \mathrm{EM}[\delta \widehat{\rho}_{L, \kappa}] = - \frac{\partial}{\partial (m L)}
\log{\delta \widehat{\rho}_{L, \kappa}} \,,
\label{eq:eff_mass}
\end{equation}
which in practice we evaluate numerically.

The effective mass analysis highlights the common asymptotic behaviour shared
among the three form-factor parametrizations and across the different values
of $\tau$. This is consistent with the expected $\frac{1}{L} e^{-m L}$ scaling,
which we derive in App.~\ref{app:B-term-subdominant}. The
deviations observed for small values of $L$ can be attributed to the
power-like prefactor on the exponential and potentially to higher order terms.

We next evaluate Eq.~\eqref{eq:final-result}, for the more interesting Cauchy
kernel
\begin{equation}
\widehat{\kappa}(\omega, \boldsymbol{\alpha}) =
\frac{1}{\pi}\frac{\sigma}{(\omega - \omega^*)^2 + \sigma^2} \,,
\label{eq:cauchy_kernel}
\end{equation}
relevant for example in the study of the smeared
$R$-ratio~\cite{Poggio:1975af}.
We restrict our analysis to the Gounaris-Sakurai model and to the target energy
$\omega^* = 3m$, and we vary the smearing radius $\sigma$ over different values
in the range $[0.5m, 2m]$.

First, in Fig.~\ref{fig:convergence_cauchy}, we investigate the convergence of
the series over Poisson modes, a good indicator of the applicability of our
result. As expected, the series is well-behaved for sufficiently large values
of $\sigma L$, where our estimate of finite-volume effects might be used to
correct future lattice calculations.

Second, in Fig.~\ref{fig:cauchy_fve}, we show the relative size of the
finite-volume contributions to the smeared spectral density, as defined in
Eq.~\eqref{eq:final-result} and summed up to $|\boldsymbol{n}| = 6$.
Note that the (apparent) singularities appearing in the figure are spurious
features simply induced by zero crossings of $\delta \widehat{\rho}_{L,\kappa}$
at specific values of $L$.
This oscillatory behaviour is a common feature in the finite-volume effects of
smeared spectral densities, also observed in
Refs.~\cite{Hansen:2017mnd,Bulava:2021fre}.

With the form factor set to unity, the contribution of the Cauchy-kernel
poles in the upper half of the complex $p$-plane can be evaluated
analytically, provided that they are the singularities closest to the real
axis (see again Fig.~\ref{fig:p-form-factor}). For the leading shell,
$|\boldsymbol{n}| = 1$, one finds
\begin{equation}\label{eq:poles_ni}
\delta\widehat{\rho}^{\,(\mathrm{ni})}_{L,\kappa} =
\frac{|p_\kappa|^2}{\pi^2 L\,|\omega_\kappa|^3}\,e^{-m_{\mathrm{env}}L}\,
\sin\big(L\,\mathrm{Re}\,[p_\kappa] + \varphi_{\kappa}\big),
\end{equation}
where
\begin{align}
\omega_\kappa &= \omega^* + i\sigma \,,
\\
p_\kappa &= \tfrac{1}{2}\sqrt{\omega_\kappa^2 - 4m^2} \,,
\\
\varphi_{\kappa} &= \arg\big(p_\kappa^2/\omega_\kappa^3\big) \,,
\\
m_{\mathrm{env}} &= \mathrm{Im}[p_\kappa] =
\tfrac{1}{2}\,\mathrm{Im}\sqrt{(\omega^* + i\sigma)^2 - 4m^2} \,.
\end{align}
The finite-volume effects thus oscillate in $L$ with frequency
$\mathrm{Re}[p_\kappa]$, modulated by an envelope that decays as
$\exp(-m_{\mathrm{env}}L)$. Up to an overall normalization, this envelope
captures the asymptotic scaling of the full result, and it is shown as the
dashed lines in Fig.~\ref{fig:cauchy_fve}. The analytic result extends
naturally to higher shells, which are further suppressed as
$\exp(-|\boldsymbol{n}|\,m_{\mathrm{env}}L)$, as well as to a non-trivial form
factor. The latter enters only through the factor $F(\omega_\kappa^2)^2$,
which modifies the amplitude and phase of the oscillation while leaving the
decay rate of the envelope unchanged.

Equation~\eqref{eq:poles_ni} also accounts for the $\sigma$ dependence seen
in Fig.~\ref{fig:cauchy_fve}. At fixed $\omega^*$, increasing $\sigma$
moves the Cauchy-kernel poles deeper into the complex plane, so that
$m_{\mathrm{env}}$ grows (from $0.33m$ at $\sigma = 0.5m$ to $0.63m$ at
$\sigma = m$, for $\omega^* = 3m$) and their contribution is increasingly
suppressed. As long as these poles remain the singularities closest to the
real axis, however, they govern the finite-volume effects and the oscillatory
pattern of Eq.~\eqref{eq:poles_ni} persists. For $\omega^* = 3m$ this
ceases to be the case once $m_{\mathrm{env}} > m$, i.e.~for $\sigma \gtrsim
1.7m$, beyond which the nearest singularity is the kinematic one at $p = im$
discussed in Sec.~\ref{sec:derivation}. The curve for $\sigma = 2m$ in
Fig.~\ref{fig:cauchy_fve} illustrates this regime: the oscillatory structure
is lost and the curve deforms into a straight line, with asymptotic slope
saturating at the standard pion-mass behaviour, $e^{-mL}$.

\section{Conclusions}
\label{sec:conclusions}

In this work we have derived a representation for the leading finite-volume
effects in smeared spectral densities, focusing on the vector-vector spectral
density relevant for the inclusive hadronic contribution to the $R$-ratio.

The result expresses the dominant finite-volume correction in terms of the
infinite-volume pion form factor evaluated at timelike kinematics, together
with the chosen smearing kernel. We have obtained the same expression in two
complementary ways: by applying the spectral-reconstruction kernel directly to
the finite-volume effects in the Euclidean correlator, and by starting from the
Lellouch-L\"uscher-Meyer representation. This agreement gives a direct
connection between the spacelike description of finite-volume effects and the
finite-volume matrix-element formalism. In Sec.~\ref{sec:numerics}, we have
additionally explored the size and structure of the corrections numerically for
representative form-factor models and smearing kernels.

The connection to the Lellouch-L\"uscher framework suggests a broad range of
future applications. In particular, the strategy introduced here should make it
possible to derive analogous finite-volume representations for any spectral
density whose corresponding finite-volume matrix elements can be connected to
infinite-volume amplitudes by a known Lellouch-L\"uscher-type relation. In this
sense, the present work can be viewed as a template: the detailed form of the
final expression will change with the relevant spectrum and matrix-element
relation, but the logic of converting the spectral sum into a Poisson-mode
representation is general.

One example future application is to spectral densities entering long-distance
contributions to $D^0$-$\overline D{}^0$ mixing. The required finite-volume
formalism for multi-hadron $D$ decays (albeit neglecting four-pion states) is
given in Ref.~\cite{Hansen:2012tf}, and the framework for extracting the
relevant spectral functions and relating them to the mixing amplitudes was
recently developed in Ref.~\cite{DiCarlo:2025mnm}. That work, however, does not
give a formula for the associated finite-volume effects, and applying the
methods introduced here should make it possible to derive such formulae.

Another natural direction is to consider spectral densities dominated by
three-particle states, such as those arising from vector iso-scalar or
axial-vector currents. In this case the corresponding finite-volume
matrix-element formalism is the three-particle generalization of the
Lellouch-L\"uscher relation, developed in Ref.~\cite{Hansen:2021ofl}. The same
steps used here should then lead to a representation of the leading
finite-volume effects in terms of the relevant three-particle amplitudes and
transition matrix elements, although the resulting expressions will be more
involved due to the richer finite-volume spectrum and the integral equations
entering the three-particle formalism.

Ultimately, the main utility of these results will be in applications to
lattice Monte Carlo data, where explicit finite-volume formulae can be combined
with spectral-reconstruction methods to improve the control of the $L \to
\infty$ limit for phenomenologically relevant smeared spectral densities.

\section*{Acknowledgements}

FAB would like to acknowledge the Higgs Centre at the University of Edinburgh
for hospitality during completion of this work.
FAB is supported by the University of Milano-Bicocca's Exchange Extra EU
Programme.
MTH is supported by the UKRI Future Leaders Fellowships MR/T019956/1 and
MR/Y034201/1 and by the STFC Consolidated Grant ST/X000494/1. At the
beginning of the project, MB was supported by the national program for young researchers ``Rita Levi Montalcini''. This work was (partially) supported by ICSC - Centro Nazionale di Ricerca in High Performance Computing, Big Data and Quantum Computing, funded by European Union - NextGenerationEU.

\bibliography{biblio.bib}

\onecolumngrid
\newpage

\appendix

\section{Subdominance of the \texorpdfstring{$\delta G_L^B(\tau)$}{deltaGL-B}
term}
\label{app:B-term-subdominant}

In this appendix we show that the contribution from the $\delta G_L^B(\tau)$
term is subdominant to the leading finite-volume correction given by $\delta
G_L^A(\tau)$, defined in Eq.~\eqref{eq:deltaGL-pole-A} of the main text. In
the final part of the appendix we additionally quantify, within a simple
monopole model for the form factor, the consequences of this term for the
finite-volume effects of the smeared spectral density, see
Eq.~\eqref{eq:epsilon-B-monopole} and Table~\ref{tab:rho-A-vs-B}.
We begin by recalling the definition~\cite{Hansen:2020whp}:
\begin{equation}
\delta G_L^B(\tau) = -\frac{1}{6} \sum_{\boldsymbol n \neq \boldsymbol 0}
\frac{1}{4\pi^3 \vert \boldsymbol n \vert L }\text{Re} \left [ \int {\rm d}
p_z e^{-\vert \boldsymbol n \vert L\sqrt{m^2 + p_z^2}} \int {\rm d} k_z \frac{
e^{i (k_z + i\mu) \vert \tau \vert}(4 m^2 + (k_z + i\mu)^2 ) } {(k_z +
i\mu)^2 - 4 p_z^2} F\!\left[-(k_z + i\mu)^2\right]^2 \right ] \,,
\label{eq:deltaGL-B-def}
\end{equation}
where the sum over images and the prefactor are the same as in
Eq.~\eqref{eq:deltaGL-pole-A}.

\bigskip

\textit{Non-interacting limit} ---
Consider first the non-interacting limit, $F=1$. In this case it is useful to
write $\delta G_L^B(\tau) = \delta G_L^{B, \delta}(\tau) + \delta G_L^{B,
\sf{rest}}(\tau)$ where
\begin{align}
\delta G_L^{B, \delta}(\tau) & = -\frac{1}{6} \sum_{\boldsymbol n \neq
\boldsymbol 0} \frac{1}{4\pi^3 \vert \boldsymbol n \vert L }\text{Re} \left [
\int {\rm d} p_z e^{-\vert \boldsymbol n \vert L\sqrt{m^2 + p_z^2}} \int {\rm
d} k_z \, e^{i (k_z + i\mu) \vert \tau \vert} F\!\left[-(k_z +
i\mu)^2\right]^2 \right ] \,,
\\
\delta G_L^{B, \sf{rest}}(\tau) & = -\frac{1}{6} \sum_{\boldsymbol n \neq
\boldsymbol 0} \frac{1}{\pi^3 \vert \boldsymbol n \vert L }\text{Re} \left [
\int {\rm d} p_z e^{-\vert \boldsymbol n \vert L\sqrt{m^2 + p_z^2}} \int {\rm
d} k_z \, e^{i (k_z + i\mu) \vert \tau \vert} \frac{ m^2 + p_z^2}{(k_z +
i\mu)^2 - 4 p_z^2} F\!\left[-(k_z + i\mu)^2\right]^2 \right ] \,.
\end{align}
For $F=1$ and $\tau \neq 0$, $\delta G_L^{B, \delta}(\tau)$ vanishes in the
sense of a distribution as follows from $\int dz e^{i k z} = 2 \pi \delta(k)$.
The second term, $\delta G_L^{B, \sf{rest}}(\tau)$, vanishes in the more
straightforward sense as a convergent integral. The integrand is an analytic
function of $k_z$ in the upper half of the complex plane and falls off fast
enough as $\vert k_z \vert \to \infty$, so that the contour can be closed
without enclosing any singularities. Thus $\delta G_L^B(\tau) = 0$ for $\tau
\neq 0$ in the non-interacting limit.

\bigskip

\textit{Large-volume comparison in the interacting case} ---
Returning to the interacting case, we now show that $\delta G_L^B(\tau)$ is
subdominant to $\delta G_L^A(\tau)$ at large $L$. To make this comparison, we
first require the large-$L$ expansion of $\delta G_L^A(\tau)$. Since we are
interested in the leading large-$L$ scaling, throughout this subsection we
retain only the leading shell, $|\boldsymbol n|=1$, whose multiplicity of six
cancels the explicit factor of $1/6$: all other shells are suppressed relative
to it by at least $e^{-(\sqrt 2 - 1) mL}$, and their contributions follow from
the results below by replacing $L \to |\boldsymbol n| L$, including in the
$1/L$ prefactor, and multiplying by the multiplicity of the shell divided by
six. Starting from Eq.~\eqref{eq:deltaGL-pole-A}, we parametrize the contour
as $k_3=k_z+i\mu$. The $L\sqrt{m^2+(k_z+i\mu)^2/4}$ exponent
is dominated by $k_z+i\mu=\mathcal O(\sqrt{m/L})$.
It is therefore useful to set
\begin{equation}
k_z+i\mu=\sqrt{\frac{m}{L}}\,z\,,
\qquad
\mu=\sqrt{\frac{m}{L}}\,\bar\mu\,,
\qquad
z\in\mathbb R+i\bar\mu \,,
\label{eq:deltaGL-A-rescaling}
\end{equation}
where $\bar\mu>0$ is held fixed as $L\to\infty$. In this region,
\begin{align}
L\sqrt{m^2+\frac{(k_z+i\mu)^2}{4}}
&=
mL+\frac{z^2}{8}
-\frac{z^4}{128mL}
+\mathcal O\!\left(1/(mL)^2\right) \,,
\\
4m^2+(k_z+i\mu)^2
&=
4m^2\left[1+\frac{z^2}{4mL}\right] \,,
\\
F\!\left[-(k_z+i\mu)^2\right]^2
&=
1-\frac{2m z^2}{L}F'(0)
+\mathcal O\!\left(1/(mL)^2\right) \,,
\end{align}
where we have used $F(0)=1$. Substitution into
Eq.~\eqref{eq:deltaGL-pole-A} gives
\begin{equation}
\delta G_L^A(\tau)
=
\frac{m^2e^{-mL}}{\pi L}\,
\mathrm{Im}\int_{\mathbb R+i\bar\mu}\frac{{\rm d}z}{2\pi}\,
\frac{e^{-z^2/8+i z\tau\sqrt{m/L}}}{z}
\left[
1+\frac{z^4}{128mL}
+\frac{z^2}{4mL}
-\frac{2m z^2}{L}F'(0)
+\mathcal O\!\left(1/(mL)^2\right)
\right] \,.
\label{eq:deltaGL-A-large-L-expanded}
\end{equation}
Thus the overall factor $e^{-mL}/L$ is fixed before carrying out the
remaining dimensionless integral. The leading integral can be evaluated by
differentiating with respect to
$b=\tau\sqrt{m/L}$ and fixing the integration constant from the contour
passing above the pole:
\begin{equation}
\int_{\mathbb R+i\bar\mu}\frac{{\rm d}z}{2\pi}\,
\frac{e^{-z^2/8+i b z}}{z}
=
-\frac{i}{2}\,
\operatorname{erfc}\!\left(\sqrt{2}\,b\right) \,.
\end{equation}
We therefore obtain the leading large-$L$ result
\begin{equation}
\delta G_L^A(\tau)
=
-\frac{m^2}{2\pi L}\,e^{-mL}\,
\operatorname{erfc}\!\left(
\tau\sqrt{\frac{2m}{L}}
\right)
+\mathcal O\!\left(
me^{-mL}/L^2
\right) \,.
\label{eq:deltaGL-A-large-L}
\end{equation}
This form keeps the dependence on $\tau/\sqrt{L}$ unexpanded. For fixed
$\tau$ it reduces to
\begin{equation}
\delta G_L^A(\tau)
=
-\frac{m^2}{2\pi L}\,e^{-mL}
\left[
1-\frac{2\sqrt{2}\,m\tau}{\sqrt{\pi mL}}
+\mathcal O\!\left(1/(mL)\right)
\right] \,.
\label{eq:deltaGL-A-large-L-fixed-tau}
\end{equation}
% In particular, $\delta G_L^A(\tau)\sim e^{-mL}/L$.

Returning now to $\delta G_L^B(\tau)$, we observe that in
Eq.~\eqref{eq:deltaGL-B-def} the $L$ dependence resides entirely in the factor
$e^{-L\sqrt{m^2+p_z^2}}$, which at large $L$ localizes the $p_z$ integral to
the region $p_z=\mathcal O(\sqrt{m/L})$, while the $k_z$ integral carries no
$L$ dependence at all. It is therefore convenient to write
\begin{equation}
\delta G_L^B(\tau)
=
-\frac{1}{4\pi^3 L}\int {\rm d}p_z\, e^{-L\sqrt{m^2+p_z^2}}\,
\operatorname{Re}\Phi(p_z,\tau)\,,
\qquad
\Phi(p_z,\tau)
=
\int_{\mathbb R+i\mu} {\rm d}k_3\,
\frac{e^{ik_3\lvert\tau\rvert}\,(4m^2+k_3^2)}{k_3^2-4p_z^2}\,
F(-k_3^2)^2\,,
\label{eq:deltaGL-B-Phi}
\end{equation}
where the poles of the integrand at $k_3=\pm 2p_z$ lie a distance $\mu$ below
the contour, so that $\Phi(p_z,\tau)$ is a smooth, $L$-independent function of
$p_z$ in the vicinity of $p_z=0$, depending on $p_z$ only through $p_z^2$.
Setting $p_z=\sqrt{m/L}\,x$ and using
$L\sqrt{m^2+p_z^2}=mL+x^2/2+\mathcal O(1/(mL))$, the $p_z$ integral is Gaussian
at leading order and one finds
\begin{equation}
\delta G_L^B(\tau)
=
-\frac{e^{-mL}}{4\pi^3 L}\sqrt{\frac{2\pi m}{L}}\,
\operatorname{Re}\Phi(0,\tau)
\left[1+\mathcal O\!\left(1/(mL)\right)\right],
\qquad
\Phi(0,\tau)
=
\int_{\mathbb R+i\mu}{\rm d}k_3\,
\frac{e^{ik_3\lvert\tau\rvert}\,(4m^2+k_3^2)}{k_3^2}\,F(-k_3^2)^2\,.
\label{eq:deltaGL-B-large-L}
\end{equation}
In the non-interacting limit, $F=1$, closing the contour in the upper
half-plane shows that $\Phi(0,\tau)=0$ for $\tau\neq0$, consistent with the
exact result derived above. For an interacting form factor, by contrast,
$\Phi(0,\tau)$ is an $L$-independent, generically non-vanishing function of
$\tau$, determined by the singularities of $F(-k_3^2)^2$ in the upper
half-plane. We conclude that
\begin{equation}
\delta G_L^B(\tau)=\mathcal O\!\left(e^{-mL}/L^{3/2}\right)\,,
\qquad
\frac{\delta G_L^B(\tau)}{\delta G_L^A(\tau)}
=
\frac{\operatorname{Re}\Phi(0,\tau)}{2\pi^2 m^2}\sqrt{\frac{2\pi m}{L}}
\left[1+\mathcal O\!\left(1/\sqrt{mL}\right)\right],
\label{eq:deltaGL-B-scaling-alt}
\end{equation}
where we have used Eq.~\eqref{eq:deltaGL-A-large-L-fixed-tau}. Thus $\delta
G_L^B(\tau)$ is suppressed by a factor $1/\sqrt{mL}$ relative to $\delta
G_L^A(\tau)$.

%We stress that this conclusion relies on the observation that
%the $k_3$ integral in Eq.~\eqref{eq:deltaGL-B-def} is not localized at small
%$k_3$: an expansion of the form factor about $k_3=0$ inside the integral is
%not uniform and would incorrectly suggest a stronger suppression.
For the
monopole form factor used in Table~\ref{tab:GL-A-vs-GL-B} and in the final
subsection, $\Phi(0,\tau)$ is given by the residue at $k_3=iM$ and we have
verified numerically that Eq.~\eqref{eq:deltaGL-B-large-L} reproduces the
exact result, with corrections of relative size $\mathcal O(1/(mL))$.
Table~\ref{tab:GL-A-vs-GL-B} compares $\delta G_L^A(\tau)$ and $\delta
G_L^B(\tau)$ numerically for this form factor over a range of $L$ values.

\renewcommand{\arraystretch}{1.5}
\begin{table}
\centering
\begin{tabular}{c c c c}
\hline
\ \ \ \ $mL$ \ \ \ \ & \ \ \ \ $\delta G_L^A$ \ \ \ \ & \ \ $\delta G_L^B$ \ \
& \ \ $\delta G_L^B / \delta G_L^A$ \ \ \\
\hline
4 & $-1.51\times10^{-4}$ & $-1.01\times10^{-5}$ & 0.0667 \\
5 & $-5.99\times10^{-5}$ & $-2.64\times10^{-6}$ & 0.0442 \\
6 & $-2.21\times10^{-5}$ & $-7.38\times10^{-7}$ & 0.0333 \\
7 & $-7.96\times10^{-6}$ & $-2.15\times10^{-7}$ & 0.0270 \\
8 & $-2.82\times10^{-6}$ & $-6.46\times10^{-8}$ & 0.0229 \\
9 & $-9.94\times10^{-7}$ & $-1.99\times10^{-8}$ & 0.0200 \\
10 & $-3.49\times10^{-7}$ & $-6.24\times10^{-9}$ & 0.0179 \\
\hline
\end{tabular}
\caption{Numerical comparison of $\delta G_L^A(\tau)$ and $\delta G_L^B(\tau)$
for $m\tau=1$ and for the values of $mL$ listed in the first column, using the
simplified monopole form factor $F(s)=[1-s/M^2]^{-1}$ with $M/m = 727/137$.
The momentum integrals defining both terms are evaluated numerically over the
full real axis, with $\delta G_L^A$ and $\delta G_L^B$ both restricted to the
leading $\lvert\boldsymbol n\rvert=1$ shell.
The decreasing ratio in the final column demonstrates that $\delta G_L^B(\tau)$
is subdominant to $\delta G_L^A(\tau)$. The decrease is faster than the
asymptotic $1/\sqrt{mL}$ of Eq.~\eqref{eq:deltaGL-B-scaling-alt} because, for
these values of $mL$, $\delta G_L^A(\tau)$ is still far from its asymptotic
form: the $\mathcal O(1/\sqrt{mL})$ correction in
Eq.~\eqref{eq:deltaGL-A-large-L-fixed-tau} equals $-0.80$ at $mL=4$ and
$-0.50$ at $mL=10$. Combining the leading term of
Eq.~\eqref{eq:deltaGL-B-large-L} with Eq.~\eqref{eq:deltaGL-A-large-L-fixed-tau}
gives $0.067$ and $0.017$ for the ratio at $mL=4$ and $10$, respectively, in
agreement with the table.\label{tab:GL-A-vs-GL-B}}
\end{table}
\renewcommand{\arraystretch}{1}

\bigskip

\textit{Monopole ansatz and the \texorpdfstring{$\tau$}{tau} dependence of
\texorpdfstring{$\delta G_L^B(\tau)$}{deltaGL-B}} ---
We close this appendix by revisiting $\delta G_L^B(\tau)$ for the simplified
monopole form factor used in Table~\ref{tab:GL-A-vs-GL-B}, following
Ref.~\cite{Hansen:2020whp}. In the representation of
Eq.~\eqref{eq:deltaGL-B-def}, the $\tau$ dependence is oscillatory and the $L$
dependence is exponentially decaying, i.e.~$\delta G_L^B(\tau)$ is of the same
type as Eq.~\eqref{eq:deltaGL-pole-A} for $\delta G_L^A(\tau)$ before the Wick
rotation of Sec.~\ref{sec:derivation}. As noted in Ref.~\cite{Hansen:2020whp},
for the monopole ansatz
\begin{equation}
F(s) = \frac{1}{1 - s/M^2} \,,
\qquad
F(-k_3^2)^2 = \frac{M^4}{(M^2 + k_3^2)^2} \,,
\label{eq:monopole-ansatz}
\end{equation}
the $k_3$ integral in $\delta G_L^B(\tau)$ can be performed in closed form, see
the second term of Eq.~(2.37) of that reference. Here we review this result and
point out that it achieves, for $\delta G_L^B(\tau)$, the same rotation from
oscillatory to exponentially decaying $\tau$ dependence that was performed for
$\delta G_L^A(\tau)$ in the main text, so that the coefficients
$c_i(\boldsymbol \alpha)$ can be applied directly.

The key difference with respect to $\delta G_L^A(\tau)$ is that the $k_3$
integrand of Eq.~\eqref{eq:deltaGL-B-def} contains no square root: the $L$
dependence enters only through $e^{-\vert \boldsymbol n \vert L\sqrt{m^2+p_z^2}}$, which is independent
of $k_3$. Writing $k_3 = k_z + i \mu$ as before, the $k_3$ integrand is
\begin{equation}
g(k_3) =
\frac{e^{i k_3 \vert \tau \vert}\,(4m^2 + k_3^2)}{k_3^2 - 4 p_z^2}
\, F(-k_3^2)^2 \,,
\label{eq:deltaGL-B-integrand}
\end{equation}
and for the monopole ansatz this is a meromorphic function of $k_3$, with
simple poles at $k_3 = \pm 2 p_z$ on the real axis and double poles at $k_3 =
\pm i M$. Since $0 < \mu < 2m < M$, only the double pole at $k_3 = iM$ lies
above the integration contour $\mathbb R + i \mu$. We can now repeat the
folding of the contour into the upper half-plane shown in
Fig.~\ref{fig:Wick-rotation}. The factor $e^{i k_3 \vert \tau \vert}$,
together with the $1/k_3^4$ fall-off of the rational factors, ensures that the
arcs at infinity do not contribute and, in the absence of a branch cut, the two
sides of the folded contour cancel identically along the positive imaginary
axis, except for a small circle around $k_3 = iM$. The oscillatory integral
thus collapses to a residue. Setting $k_3 = i z$ to make the decay in $\vert
\tau \vert$ manifest, one finds
\begin{equation}
\int_{\mathbb R + i \mu} \frac{{\rm d} k_3}{2\pi} \, g(k_3)
=
i \operatorname*{Res}_{k_3 = iM} g(k_3)
=
- \frac{{\rm d}}{{\rm d} z}
\left[
\frac{e^{- z \vert \tau \vert} \, (z^2 - 4m^2) \, M^4}{(z + M)^2 \, (z^2 + 4
p_z^2)}
\right]_{z = M} \,,
\label{eq:deltaGL-B-residue}
\end{equation}
and substituting into Eq.~\eqref{eq:deltaGL-B-def} gives
\begin{equation}
\delta G_L^B(\tau)
=
\frac{1}{6} \sum_{\boldsymbol n \neq \boldsymbol 0}
\frac{1}{2\pi^2 \vert \boldsymbol n \vert L}
\int {\rm d} p_z \, e^{-\vert \boldsymbol n \vert L\sqrt{m^2 + p_z^2}} \,
\frac{{\rm d}}{{\rm d} z}
\left[
\frac{e^{- z \vert \tau \vert} \, (z^2 - 4m^2) \, M^4}{(z + M)^2 \, (z^2 + 4
p_z^2)}
\right]_{z = M} \,,
\label{eq:deltaGL-B-monopole}
\end{equation}
which is precisely the second term of Eq.~(2.37) of
Ref.~\cite{Hansen:2020whp}, now with the full sum over images. Carrying out
the derivative, the result can be written as
\begin{equation}
\delta G_L^B(\tau) = e^{-M \vert \tau \vert} \big [ a_L - b_L \vert \tau
\vert \big ] \,,
\label{eq:deltaGL-B-monopole-ab}
\end{equation}
with
\begin{align}
a_L & = \frac{1}{6} \sum_{\boldsymbol n \neq \boldsymbol 0}
\frac{M^2}{8 \pi^2 \vert \boldsymbol n \vert L} \int {\rm d} p_z \,
e^{-\vert \boldsymbol n \vert L\sqrt{m^2 + p_z^2}}
\left [ \frac{M^2 + 4 m^2}{M (M^2 + 4 p_z^2)} - \frac{2 M (M^2 - 4
m^2)}{(M^2 + 4 p_z^2)^2} \right ] \,,
\label{eq:deltaGL-B-monopole-a}
\\
b_L & = \frac{1}{6} \sum_{\boldsymbol n \neq \boldsymbol 0}
\frac{M^2}{8 \pi^2 \vert \boldsymbol n \vert L} \int {\rm d} p_z \,
e^{-\vert \boldsymbol n \vert L\sqrt{m^2 + p_z^2}}
\, \frac{M^2 - 4 m^2}{M^2 + 4 p_z^2} \,.
\label{eq:deltaGL-B-monopole-b}
\end{align}
We have checked that Eq.~\eqref{eq:deltaGL-B-monopole-ab}, restricted to the
leading shell, reproduces the values of $\delta G_L^B$ in
Table~\ref{tab:GL-A-vs-GL-B}, which were obtained by direct numerical
integration of Eq.~\eqref{eq:deltaGL-B-def}.

Two comments are in order. First, the $\tau$ dependence of
Eq.~\eqref{eq:deltaGL-B-monopole-ab} is manifestly that of a double pole at
$\omega = M$: a decaying exponential multiplied by a polynomial of degree one
in $\vert \tau \vert$. Using Eq.~\eqref{eq:GL-rhoL}, this is equivalent to a
contribution to the finite-volume spectral density of the form
\begin{equation}
\omega^2 \, \delta \rho_L^B(\omega) = a_L \, \delta(\omega - M) - b_L \,
\delta'(\omega - M) \,,
\label{eq:deltaGL-B-monopole-rho}
\end{equation}
i.e.~within the monopole ansatz, $\delta G_L^B(\tau)$ is the finite-volume
correction associated with a stable state of mass $M$. This is the imprint of
the pole of the monopole form factor at $s = M^2$, which lies on the real axis
and, since $M > 2m$, replaces the two-pion cut by a stable vector meson. For a
form factor with the analytic structure discussed in
Sec.~\ref{sec:derivation}, the corresponding singularity is a resonance pole on
the second sheet and no such isolated contribution arises. Second, for each shell
$a_L$ and $b_L$ are given by the Gaussian-like weight $e^{-\vert \boldsymbol n
\vert L\sqrt{m^2+p_z^2}}$ integrated against bounded, $L$-independent functions
of $p_z$, so that the leading shell scales as $a_L, b_L \sim e^{-mL}/L^{3/2}$
as $L \to \infty$, with coefficients that can be read off by setting $p_z = 0$
in the integrands. This is the general result of
Eq.~\eqref{eq:deltaGL-B-scaling-alt}, with $\operatorname{Re}\Phi(0,\tau)$
given here by the residue of the integrand of Eq.~\eqref{eq:deltaGL-B-large-L}
at $k_3 = iM$.

Finally, since Eq.~\eqref{eq:deltaGL-B-monopole-ab} is exponentially decaying
in $\tau$, the coefficients $c_i(\boldsymbol \alpha)$ of
Eq.~\eqref{eq:kappa-expansion} can be applied to it directly, exactly as was
done for $\delta G_L^A(\tau)$ in reaching Eq.~\eqref{eq:final-result}. In the
same approximation as in the main text, $\widehat \kappa_{\sf r} \to \widehat
\kappa$, one has $\sum_i c_i(\boldsymbol \alpha) e^{-\omega \tau_i} = \widehat
\kappa(\omega, \boldsymbol \alpha)/\omega^2$ and $\sum_i c_i(\boldsymbol
\alpha) \tau_i e^{-\omega \tau_i} = - \partial_\omega [\widehat \kappa(\omega,
\boldsymbol \alpha)/\omega^2]$, so that the contribution of $\delta
G_L^B(\tau)$ to $\epsilon_{L,\kappa}(\boldsymbol \alpha)$ is
\begin{align}
\epsilon^B_{L,\kappa}(\boldsymbol \alpha)
\equiv \sum_{i=1}^N c_i(\boldsymbol \alpha) \, \delta G_L^B(\tau_i)
& =
a_L \, \frac{\widehat \kappa(M, \boldsymbol \alpha)}{M^2}
+ b_L \, \frac{\partial}{\partial \omega} \left [ \frac{\widehat
\kappa(\omega, \boldsymbol \alpha)}{\omega^2} \right ]_{\omega = M} \,,
\label{eq:epsilon-B-monopole}
\\
& =
\frac{1}{6} \sum_{\boldsymbol n \neq \boldsymbol 0}
\frac{1}{2\pi^2 \vert \boldsymbol n \vert L}
\int {\rm d} p_z \, e^{-\vert \boldsymbol n \vert L\sqrt{m^2 + p_z^2}} \,
\frac{{\rm d}}{{\rm d} z}
\left[
\frac{(z^2 - 4m^2) \, M^4}{(z + M)^2 \, (z^2 + 4 p_z^2)} \,
\frac{\widehat \kappa(z, \boldsymbol \alpha)}{z^2}
\right]_{z = M} \,,
\label{eq:epsilon-B-monopole-2}
\end{align}
where the second form follows directly from Eq.~\eqref{eq:deltaGL-B-monopole}.
Thus, within the monopole ansatz, the $\delta G_L^B(\tau)$ contribution to the
finite-volume effects of the smeared spectral density is determined by the
smearing kernel and its first derivative at $\omega = M$. We stress that this is a statement within the
model: the monopole ansatz is used here only to make the analytic structure of
$\delta G_L^B(\tau)$ explicit, and Eq.~\eqref{eq:epsilon-B-monopole} should
not be read as a quantitative estimate of $\epsilon_{L,\kappa}(\boldsymbol
\alpha)$ for a realistic form factor.

\renewcommand{\arraystretch}{1.5}
\begin{table}
\centering
\begin{tabular}{c c c c c}
\hline
\ \ $\sigma/m$ \ \ & \ \ $mL$ \ \ & \ \ $f_\kappa(L,\boldsymbol\alpha)$ \ \ & \ \
$\epsilon_{L,\kappa}^B(\boldsymbol\alpha)$ \ \ & \ \ $\epsilon_{L,\kappa}^B / f_\kappa$ \ \ \\
\hline
& 4 & $-4.40\times10^{-4}$ & $-3.11\times10^{-6}$ & $0.00708$ \\
 & 5 & $1.93\times10^{-4}$ & $-7.49\times10^{-7}$ & $-0.00389$ \\
$0.5$ & 6 & $1.05\times10^{-4}$ & $-1.98\times10^{-7}$ & $-0.00189$ \\
 & 7 & $8.39\times10^{-5}$ & $-5.57\times10^{-8}$ & $-0.000664$ \\
 & 8 & $1.14\times10^{-5}$ & $-1.64\times10^{-8}$ & $-0.00143$ \\
\hline
 & 4 & $-1.61\times10^{-4}$ & $-4.47\times10^{-6}$ & $0.0278$ \\
 & 5 & $6.15\times10^{-5}$ & $-1.07\times10^{-6}$ & $-0.0175$ \\
$0.8$ & 6 & $4.91\times10^{-5}$ & $-2.84\times10^{-7}$ & $-0.00579$ \\
 & 7 & $1.94\times10^{-5}$ & $-7.98\times10^{-8}$ & $-0.00411$ \\
 & 8 & $-7.71\times10^{-6}$ & $-2.35\times10^{-8}$ & $0.00304$ \\
\hline
 & 4 & $-9.84\times10^{-5}$ & $-5.08\times10^{-6}$ & $0.0516$ \\
 & 5 & $2.97\times10^{-5}$ & $-1.22\times10^{-6}$ & $-0.0411$ \\
$1.0$ & 6 & $2.48\times10^{-5}$ & $-3.23\times10^{-7}$ & $-0.013$ \\
 & 7 & $5.35\times10^{-6}$ & $-9.08\times10^{-8}$ & $-0.017$ \\
 & 8 & $-5.04\times10^{-6}$ & $-2.67\times10^{-8}$ & $0.0053$ \\
\hline
\end{tabular}
\caption{Finite-volume effects on the Cauchy-smeared spectral density, in
units $m = 1$, for $\omega^* = 3m$ and $\sigma/m = 0.5$, $0.8$ and $1.0$, three
of the smearing widths used in Fig.~\ref{fig:cauchy_fve}, with
$f_\kappa(L,\boldsymbol\alpha)$ and
$\epsilon^B_{L,\kappa}(\boldsymbol\alpha)$ both summed over images with
$\vert\boldsymbol n\vert \leq 6$, as in Fig.~\ref{fig:cauchy_fve}. The
column $f_\kappa(L,\boldsymbol\alpha)$ is the main result,
Eq.~\eqref{eq:final-result}, evaluated with the Gounaris-Sakurai form factor
and the parameters of Sec.~\ref{sec:numerics}, while
$\epsilon^B_{L,\kappa}(\boldsymbol\alpha)$ is given by
Eq.~\eqref{eq:epsilon-B-monopole} with the monopole form factor of
Table~\ref{tab:GL-A-vs-GL-B}, $M/m = 727/137$. The two columns therefore rely
on different form-factor models, see the text. The
$f_\kappa(L,\boldsymbol\alpha)$ entries are consistent with
Fig.~\ref{fig:cauchy_fve}, but are not normalized to the infinite-volume
smeared spectral density $\widehat\rho_\kappa$ as they are there.
%The entries can be related to
% those of Table~\ref{tab:GL-A-vs-GL-B} by replacing the Cauchy kernel with the
% exponential kernel $\widehat\kappa(\omega, \boldsymbol\alpha) = \omega^2
% e^{-\omega\tau}$ at $m\tau = 1$ and restricting to the leading shell:
% Eq.~\eqref{eq:epsilon-B-monopole} then reduces
% to $\delta G_L^B(\tau)$, Eq.~\eqref{eq:deltaGL-B-monopole-ab}, and reproduces
% the $\delta G_L^B$ column of Table~\ref{tab:GL-A-vs-GL-B}, while
% Eq.~\eqref{eq:final-result} reduces to the leading-shell term of $\delta
% G_L(\tau)$ in Eq.~\eqref{eq:deltaGL-final}.
The sign changes of
$f_\kappa(L,\boldsymbol\alpha)$ reflect the oscillatory $L$ dependence visible
in Fig.~\ref{fig:cauchy_fve}, so that the ratio in the last column is not
monotonic in $L$.\label{tab:rho-A-vs-B}}
\end{table}
\renewcommand{\arraystretch}{1}

To give an indication of the size of this contribution relative to the leading
term, Table~\ref{tab:rho-A-vs-B} lists, for the Cauchy kernel with $\omega^* =
3m$ and $\sigma/m = 0.5$, $0.8$ and $1.0$, the
values of $f_\kappa(L, \boldsymbol \alpha)$ from Eq.~\eqref{eq:final-result},
summed over images with $\vert \boldsymbol n \vert \leq 6$ as in Fig.~\ref{fig:cauchy_fve}
and evaluated with the Gounaris-Sakurai form factor of
Sec.~\ref{sec:numerics},
alongside $\epsilon^B_{L,\kappa}$ from Eq.~\eqref{eq:epsilon-B-monopole} with
the monopole form factor of Table~\ref{tab:GL-A-vs-GL-B}. Since no closed form
for $\delta G_L^B(\tau)$ is available for the Gounaris-Sakurai model, the two
columns necessarily use different form factors, and the comparison is
indicative only. With this caveat, away from the zero crossings of
$f_\kappa(L, \boldsymbol \alpha)$ the $B$ term is at or below the few-percent
level for all three values of $\sigma$.

\section{Proof that \texorpdfstring{$\delta G_L^{A,-}(\tau) = \delta
G_L^{A,+}(\tau)$}{deltaGL-A-minus equals
deltaGL-A-plus}\label{app:minus-equals-plus}}

In this appendix we show that the negative branch of the Wick-rotated contour
gives an identical contribution to the positive branch, i.e.~$\delta
G_L^{A,-}(\tau) = \delta G_L^{A,+}(\tau)$.

Starting from the definition,
\begin{equation}
\delta G_L^{A,-}(\tau) = \frac{1}{6} \sum_{\boldsymbol n \neq \boldsymbol 0}
\mathrm{Im} \!\left[
e^{-i\theta}
\int_{-\infty}^{0}
\frac{\mathrm{d}x}{2\pi}
\frac{
e^{\, i (x e^{-i\theta} + i\mu)\vert \tau \vert
- \vert \boldsymbol n \vert L \sqrt{m^2 + (x e^{-i\theta} + i\mu)^2/4}}
}{
4\pi (x e^{-i\theta} + i\mu) \vert \boldsymbol n \vert L
}
\left(
4m^2 + (x e^{-i\theta} + i\mu)^2
\right)
F\!\left[-(x e^{-i\theta} + i\mu)^2\right]^2
\right] \,,
\end{equation}
we first perform the change of variables $x \to -x$ and complex conjugate the
integrand while adding a minus outside the integral to compensate. This gives
\begin{multline}
\delta G_L^{A,-}(\tau) = - \frac{1}{6} \sum_{\boldsymbol n \neq \boldsymbol 0}
\mathrm{Im} \!\Bigg[
e^{i\theta}
\int_{0}^{\infty}
\frac{\mathrm{d}x}{2\pi}
\frac{
e^{\, -i (-x e^{i\theta} - i\mu)\vert \tau \vert
}
}{
4\pi (-x e^{i\theta} - i\mu) \vert \boldsymbol n \vert L
}
\left(
4m^2 + (-x e^{i\theta} - i\mu)^2
\right)
\\
\times
\left (
e^{
- \vert \boldsymbol n \vert L \sqrt{m^2 + (-x e^{-i\theta} + i\mu)^2/4}}
F\!\left[-(-x e^{-i\theta} + i\mu)^2\right]^{2}
\right )^*
\Bigg] \,.
\label{eq:deltaGL-pole-minus-1}
\end{multline}

Careful consideration is needed to ensure that the square-root functions, both
explicit and implicit in the definition of $F$, are treated consistently. The
essential idea is that we are using the principal branch of the square root,
defined with a branch cut along the negative real axis. Thus, for any complex
number $z = r e^{i \phi}$ with $r \geq 0$ and $\phi \in (-\pi, \pi]$, we have
$\sqrt{z} = \sqrt{r} e^{i \phi/2}$. Excluding $\phi = \pi$ this implies
$\sqrt{z^*} = (\sqrt{z})^*$ as well as $F(-z^2)^* = F(- (z^*)^2)$. From
Eq.~\eqref{eq:deltaGL-pole-minus-1} then immediately follows that $\delta
G_L^{A,-}(\tau) = \delta G_L^{A,+}(\tau)$.

\section{Properties of \texorpdfstring{$\phi(q)$}{the pseudo-phase}}
\label{app:phi_expansion_and_positivity}

In this appendix we derive Eq.~\eqref{eq:QC_expansion} of the main text, which
gives the large-$L$ expansion of $e^{2 i \phi(q)}$. We also discuss key
properties of $\phi(q)$ itself that are relevant for the arguments in
Sec.~\ref{sec:LL}, specifically that $\mathcal Q(E,L) = \delta_{\pi\pi}(p(E)) +
\phi(q(E,L)) > 0$ for $E> 2m$ and that $n=1$ in Eq.~\eqref{eq:Q-based-QC}
corresponds to the ground state level in the finite volume. In this work the
definition of the isolated function, i.e.~not in the exponential form, is only
required for real $p$ to demonstrate these two claims, while in general all
results assume that $p$, the back-to-back momentum of the two pions, is complex
with $\mathrm{Im}[p] = \mu >0$ unless otherwise stated.\\

\textit{Large volume expansion of $e^{2 i \phi(q)}$} ---
Writing $q=pL/(2\pi)$, we recall the definition
\begin{equation}
\cot \phi(q) =
 - \frac{1}{ \pi p L } \lim_{\Lambda \to \infty}
\bigg[ \sum_{\boldsymbol n}^\Lambda \frac{1}{ \boldsymbol n^2-(pL/(2 \pi))^2 }
- 4\pi \Lambda \bigg] \,.
\label{eq:cot-phi-def}
\end{equation}
We begin by changing the summed coordinate from $\boldsymbol n$ to $\boldsymbol
k = 2 \pi \boldsymbol n / L$:
\begin{align}
\cot \phi(q)& = - \frac{4 \pi}{ p } \lim_{\overline \Lambda \to \infty}
\bigg[ \frac{1}{L^3} \sum_{\boldsymbol k}^{\overline \Lambda} \frac{1}{
\boldsymbol{k}^2 - p^2 } - \frac{\overline \Lambda}{2 \pi^2} \bigg] \,,
\end{align}
with $\overline \Lambda = 2 \pi \Lambda/L$. Using Poisson's summation formula
then gives
\begin{align}
\cot \phi(q) & = -
\frac{4 \pi}{ p } \lim_{\overline\Lambda \to \infty}
\bigg[ \frac{1}{L^3} \sum_{\boldsymbol k}^{\overline \Lambda} \int d^3
\boldsymbol x \, \frac{\delta^3(\boldsymbol{x} - \boldsymbol{k})}{
\boldsymbol{x}^2 - p^2 } - \frac{\overline \Lambda}{2 \pi^2} \bigg] \,, \\
& = - \frac{4 \pi}{ p } \int \frac{ d^3 \boldsymbol x}{(2 \pi)^3} \,
\bigg[\frac{1}{ \boldsymbol{x}^2 - p^2 } - \frac{1}{\boldsymbol{x}^2}\bigg]-
\frac{4 \pi}{ p } \sum_{\boldsymbol n \neq \boldsymbol 0} \int \frac{ d^3
\boldsymbol x}{(2 \pi)^3} \,
\frac{ e^{i L \boldsymbol n \cdot \boldsymbol x}}{ \boldsymbol{x}^2 - p^2 } \,,
\end{align}
where in the last line we have separated the contribution of $\boldsymbol n =
\boldsymbol 0$ from the rest.
In the second line we rewrite $\overline \Lambda/(2\pi^2)$ by the integral over
$1/\boldsymbol x^2$, and then send $\overline \Lambda \to \infty$, taking
advantage of the fact that the difference between the two integrands scales as
$1/\boldsymbol x^4$ and thus the integral is convergent.

This first integral can be evaluated directly to reach
\begin{equation}
 \frac{4 \pi}{ p } \int \frac{ d^3 \boldsymbol x}{(2 \pi)^3} \,
\bigg[\frac{1}{ \boldsymbol{x}^2 - p^2 } - \frac{1}{\boldsymbol{x}^2}\bigg]
= \frac{2}{ \pi } \int_0^\infty dx \, \frac{p}{x^2-p^2} = \frac{1}{ \pi }
\int_{-\infty}^\infty dx \, \frac{p}{x^2-p^2}
= i \,.
\end{equation}
In the penultimate equality we used the evenness of the integrand to extend the
integral over the full real axis. Closing this contour in either the upper or
lower half-plane gives the residue of the encircled pole at $x= \pm p$ (where
we recall that $\mathrm{Im}[p]>0$), which yields the result $i$.

For the terms with
$\boldsymbol n \neq \boldsymbol 0$ we instead obtain
\begin{equation}
\frac{4 \pi}{p} \sum_{\boldsymbol n \neq \boldsymbol 0} \int \frac{ d^3
\boldsymbol x}{(2 \pi)^3} \,
\frac{ e^{i L \boldsymbol n \cdot \boldsymbol x}}{ \boldsymbol{x}^2 - p^2 } =
\frac{1}{2 \pi i L p} \sum_{\boldsymbol n \neq \boldsymbol 0} \frac{1}{ \vert
\boldsymbol n \vert }
\int_{- \infty}^\infty \! \! dx \, x \, \frac{e^{i L \vert \boldsymbol n \vert
x} - e^{-i L \vert \boldsymbol n \vert x}}{(x - p)(x + p) } =
\frac{1}{Lp} \sum_{\boldsymbol n \neq \boldsymbol 0} \frac{1}{ \vert
\boldsymbol n \vert } e^{i L \vert \boldsymbol n \vert p} \,.
\end{equation}
In the final step, the term with $e^{i L \vert \boldsymbol n \vert x}$ is
evaluated by closing the contour in the upper half-plane, and that with $e^{-i
L \vert \boldsymbol n \vert x}$ in the lower half-plane. This gives the same
contribution after accounting for the relative minus sign between the two
exponentials. We arrive at the following expression for $\cot \phi(q)$:
\begin{equation}
 \cot \phi(q) = -i
- \frac{\zeta(pL)}{pL} \,, \qquad\qquad \zeta(z) = \sum_{\boldsymbol n \neq
\boldsymbol 0} \frac{e^{i z \vert \boldsymbol n \vert}}{\vert \boldsymbol n
\vert }
= \sum_{n > 0} \nu_n \frac{e^{i \sqrt{n} z}}{\sqrt{n}} \,,
\end{equation}
where $\nu_n$ denotes the number of integer vectors $\boldsymbol n$ with
$\boldsymbol n^2 = n$.

We next use
\begin{equation}
- i \cot z = \frac{e^{2iz}+1}{e^{2iz}-1} \ \ \ \Longrightarrow \ \ \ e^{2 i z}
= \frac{ \cot z+i}{ \cot z-i} \,,
\end{equation}
to conclude
\begin{align}
e^{2i \phi(q)} =\frac{i+\zeta(pL)/(pL)-i}{i+\zeta(pL)/(pL) +i}
= \frac{1}{2ipL}\frac{\zeta(pL)}{1+\zeta(pL)/(2i p L)} \,.
\end{align}
Expanding the denominator, we deduce Eq.~\eqref{eq:QC_expansion} of the main
text
\begin{equation}
e^{2 i \phi(q)}
= \frac{1}{2 ipL} \sum_{\boldsymbol n \neq \boldsymbol 0} \frac{e^{i \vert
\boldsymbol n \vert pL} }{\vert \boldsymbol n \vert}
+ \mathcal O(e^{{2}ipL }/(Lp)^2) \,.
\end{equation}

\begin{figure}[t]
\includegraphics[width=0.9\textwidth]{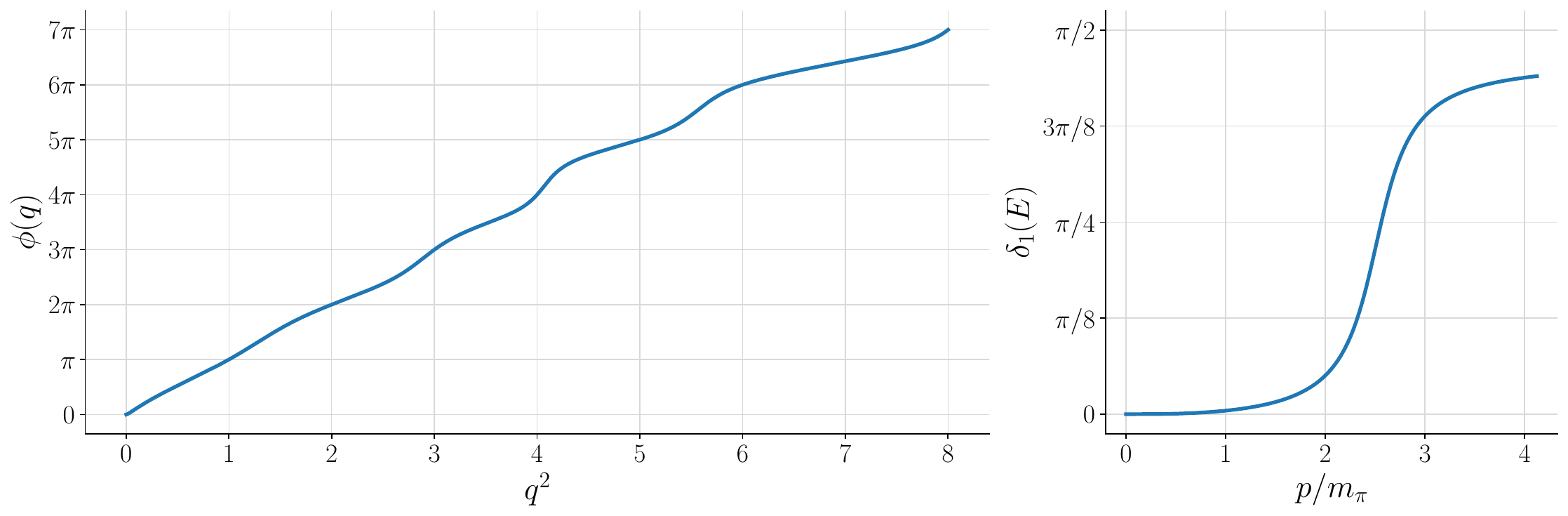}
\caption{\emph{Left:} The L\"uscher pseudo-phase $\phi(q)$,
defined by the continuous branch of the inverse cotangent, shown as a function
of $q^2$. \emph{Right:} The vector-isovector scattering phase shift
$\delta_{\pi\pi}(p(E))$ determined from the Gounaris-Sakurai model.}
\label{fig:phi-q-vs-q-squared}
\end{figure}

\textit{Large volume expansion of $\phi(q)$} ---
Having completed the essential arguments, we now continue the discussion of the
behaviour of $\phi(q)$ for complex $p$ with $\mathrm{Im}[p]>0$, since
this provides an interesting alternative perspective on the large-$L$ expansion
of $e^{2 i \phi(q)}$ and on methods for predicting the finite-volume
spectrum at large $L$. From the above results we have
\begin{equation}
\phi(q) = - \cot^{-1} \bigg(i + \frac{\zeta(pL)}{pL} \bigg) =
\frac{1}{2i} \log \bigg(\frac{2i + \zeta(pL)/pL}{\zeta(pL)/pL} \bigg) \,.
\end{equation}
Here $\cot^{-1}$ and $\log$ are both taken on the continuous branch connected
to the real-valued pseudo-phase used in the finite-volume quantization
condition of Sec.~\ref{sec:LL}. This fixes the otherwise ambiguous additive
multiples of $\pi$.
The corresponding behaviour for real $q^2$ is shown in
Fig.~\ref{fig:phi-q-vs-q-squared} where we also note that the convention
requires that the curve goes through the origin.

\begin{figure}[t]
\includegraphics[width=\textwidth]{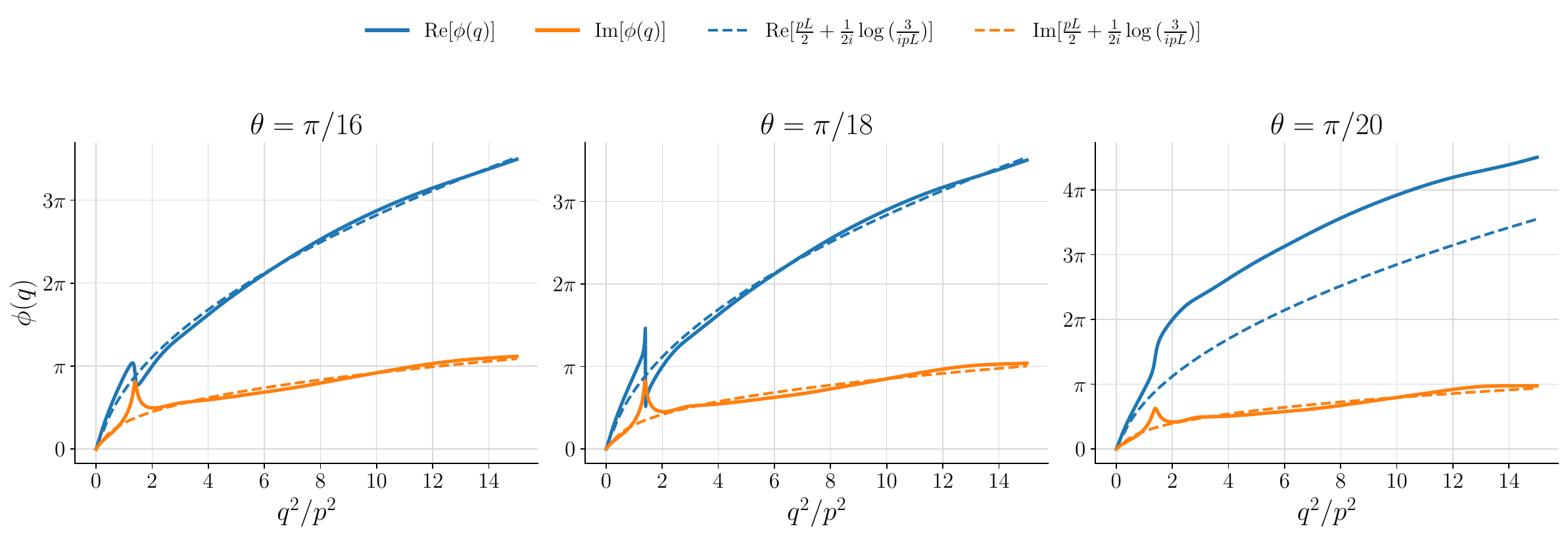}
\caption{Comparison of the L\"uscher pseudo-phase $\phi(q)$,
evaluated along complex momenta $q = \vert q \vert e^{i \theta}$, with the
leading large-volume expansion, Eq.~\eqref{eq:fpL_final}.
Here we interpret the complex $q$ trajectory as fixed complex $p = m e^{i
\theta}$ such that $q^2/p^2 = L^2 m^2 / (4 \pi^2)$ is real. The series of
panels exhibits an interesting feature. For $\theta = \pi/16$ the expansion
gives a good approximation for the large-$L$ behaviour. As $\theta$ decreases to
$\pi/18$, the function becomes rapidly varying around $q^2/p^2 = 1$. This rapid
variation then becomes a true discontinuity and the prescription to keep
$\phi(q)$ continuous leads to an additive shift, such that the large $L$
expansion for $\theta = \pi/20$ is off by $\pi$. As $\theta$ continues to
decrease towards zero at fixed $\vert p \vert$, these jumps become dense so
that the expansion cannot be applied directly for real $p$, as also expected
from the fact that the exponentials become oscillatory rather than decaying.
}
\label{fig:complex-theta-phi-expansion}
\end{figure}

By contrast, the behaviour of the complex pseudo-phase and its large-$L$ expansion
is illustrated in Fig.~\ref{fig:complex-theta-phi-expansion}.
For fixed $p$ with $\mathrm{Im}[p]>0$, the leading shell has $\vert \boldsymbol
n \vert=1$ and multiplicity six. Thus
\begin{equation}
\phi(q) = \frac{1}{2i}
 \log \left [ \frac{3}{ipL} e^{ipL}
\bigg(1+\mathcal O(e^{i(\sqrt{2}-1)pL}) \bigg) \right ] \,.
\end{equation}
Expanding the result gives
\begin{equation}
\phi(q)
= \frac{pL}{2} + n(p) \pi + \frac{1}{2 i } \log\bigg(\frac{3}{ipL}\bigg)
+ \mathcal O(e^{i(\sqrt{2}-1)pL}) \,,
\label{eq:fpL_final}
\end{equation}
where $n(p) \pi$ reflects the ambiguity in the large $L$ branch needed to
ensure continuity. The value required depends on $p$ as explained in the
caption of Fig.~\ref{fig:complex-theta-phi-expansion}.
Eq.~\eqref{eq:fpL_final} shows that the L\"uscher kinematic function grows
linearly with $L$ at fixed energy. This is the scaling used in
Sec.~\ref{sec:LL} to identify the large-volume indexing of the finite-volume
levels.

\bigskip

\textit{Positivity of $\mathcal Q(E,L)$} ---
Our convention on $\cot^{-1}$ of the expression in
Eq.~\eqref{eq:cot-phi-def}, i.e.\ requiring that $\phi(q(E,L))$ is continuous
in $E$ and $L$ for $E>2m$ and $L>0$ (or equivalently, that $\phi(q)$ is
continuous in $q^2$ for $q^2 > 0$), together with the fact that the $\pi \pi$
scattering phase shift is positive in the channel considered here, ensures that
$\mathcal Q(E,L) = \delta_{\pi\pi}(p(E)) + \phi(q(E,L)) > 0$. This, in turn,
justifies extending the sum over $n$ in Eq.~\eqref{eq:rho-hat-LL-3}, as the
non-positive-$n$ terms do not contribute.

\bigskip

\textit{Correspondence between the ground state and $n=1$} ---
Next we take $n=1$ in Eq.~\eqref{eq:Q-based-QC} and observe this corresponds to
the point where $\delta_{\pi\pi}(p(E)) = \pi - \phi(q(E,L))$. The right-hand
side vanishes for $q^2=1 \ \Rightarrow \ p = 2 \pi /L$. For large $L$ this
corresponds to a near-threshold energy, for which
\begin{equation}
\delta_{\pi\pi}(p) = a_1 p^3 + \mathcal O(p^5) \,,
\end{equation}
where $a_1$ is the $p$-wave scattering volume. Using
$\phi'(1) = 4\pi^2/6$, the solution to the quantization condition at leading
order in the large-$L$ expansion is given by
\begin{equation}
\left(\frac{Lp}{2\pi}-1\right)\phi'(1)
= -a_1 p^3
\quad \Longrightarrow \quad
p = \frac{2\pi}{L}
- \frac{24\pi^2a_1}{L^4}
+ \mathcal O(1/L^6) \,.
\end{equation}
Relating $p$ to $E_1(L)$ by $E_1(L) = 2 \sqrt{p^2 + m^2}$ and expanding to
leading order in $a_1$, we deduce~\cite{Luscher:1986pf}
\begin{equation}
E_1(L)
= 2\sqrt{(2\pi/L)^2+m^2}
- \frac{96\pi^3a_1}{L^5\sqrt{(2\pi/L)^2+m^2}}
+ \mathcal O(1/L^7) \,.
\end{equation}
This is the expected large-volume behaviour of the ground state level in the
finite volume. This final observation also requires considering the set of non-interacting
two-pion finite-volume states and projecting onto the subset with the quantum
numbers of interest, namely isospin 1 and the finite-volume irreducible
representation $T_1^-$. The latter exercise reveals that the non-interacting
ground-state level is $E_1(L) = 2 \sqrt{(2 \pi /L)^2 + m^2}$.

\end{document}